\documentstyle[12pt]{article}

\newdimen\tdim

\catcode`\@=11

\def\@maketitle{\newpage   
 \null
 \vspace*{-1\headsep}      
 \vspace*{-1\headheight}
 \vspace*{-24pt}
 \begin{flushright}{\large       
   { \preprintno} \\ \@date}
 \end{flushright}
 \vskip \headsep     
 \vskip \headheight
 \bigskip
 \begin{center}            

{\LARGE \@title \par}
   \vskip 2em
   {\large
     \lineskip .5em
     \begin{tabular}[t]{c}\@author
     \end{tabular}\par}
   \vskip 1em
 \end{center}
 \par
 \vskip 1.5em}

\newcommand{\preprintno}{preprint number here}   

\def\abstract{\if@twocolumn
\section*{Abstract}
\else                         
\begin{center}
{\bf Abstract\vspace{-.5em}\vspace{0pt}}
\end{center}
\quotation
\fi}
\def\endabstract{\if@twocolumn\else\endquotation\fi}

\def\appendix{\par
    \setcounter{section}{0}
    \setcounter{subsection}{0}
    \renewcommand{\theequation}{\Alph{section}.\arabic{equation}}
    \setcounter{equation}{0}
}

\@addtoreset{equation}{section}
\def\theequation{\arabic{section}.\arabic{equation}}

\def\ps@columns{%
 \if@twocolumn
  \let\@mkboth\@gobbletwo
  \def\@oddhead{}\def\@evenhead{}
  \def\@oddfoot%
   {\rm\hfil\thepage\stepcounter{page}\hskip.5\textwidth\thepage\hfil}
  \let\@evenfoot\@oddfoot
 \else
 \ps@plain
 \fi
}

\catcode`\@=12

\makeatletter \input art10.sty \makeatother
\def\starttext{\twocolumn}

\typein{*** or hit [return] for landscape mode(11x8.5) ***}

\newcommand{\beq}{\begin{equation}}
\newcommand{\eeq}{\end{equation}}

\newcommand{\remove}[1]{}
\newcommand{\rep}[1]{{\bf{#1}}}
\newcommand{\su}[1]{SU(${#1}$)}

\renewcommand{\theequation}{\thesection.\arabic{equation}}
\def\MeV{\rm MeV}
\def\pred{\rm pred}
\def\xp{\rm exp}
\def\i{\rm i}
\def\tot{\rm tot}

\def\fit{\rm fit}
\begin{document}

\title{Decays of $\ell=1$ Baryons --- Quark Model versus Large-$N_c$}

\author{
        Christopher D. Carone\thanks {carone@huhepl.harvard.edu} \\
        Howard Georgi \thanks{georgi@huhepl.harvard.edu} \\
        Lev Kaplan \thanks {kaplan@huhepl.harvard.edu} \\
        David Morin \thanks {morin@huhepl.harvard.edu} \\
        Lyman Laboratory of Physics \\
        Harvard University \\
        Cambridge, MA 02138}
\date{\today}

\renewcommand{\preprintno}{HUTP-94/A008}

\begin{titlepage}

\maketitle

\def\thepage {}        

\begin{abstract}
We study nonleptonic decays of the orbitally excited, \su6 \rep{70}-plet
baryons in order to test the hypothesis that the successes of the
nonrelativistic quark model have a natural explanation in the large-$N_c$
limit of QCD.  By working in a Hartree approximation, we isolate a specific
set of operators that contribute to the observed s- and d-wave decays in
leading order in $1/N_c$. We fit our results to the current experimental decay
data, and make predictions for a number of allowed but unobserved modes. Our
tentative conclusion is that there is more to the nonrelativistic quark model
of baryons than large-$N_c$.
\end{abstract}

\end{titlepage}

\starttext 
\pagestyle{columns} 
\pagenumbering {arabic} 

\section {Introduction} \label {sec:intro}

In the nonrelativistic quark model (NRQM), the baryon resonances can
be classified by their transformation properties under
nonrelativistic \su6 spin-flavor symmetry.  The ground-state
baryons have completely symmetric spin-flavor wavefunctions,
and form the 56-dimensional representation.  The $l=1$ orbitally
excited states have spin-flavor wavefunctions with mixed
symmetry that lie in the \rep{70}.  While the NRQM description of the
baryon states has not been derived convincingly from QCD, it has been
incorporated with some success in many of the previous theoretical attempts
to understand the observed baryon masses and decay widths ~\cite{isgur}.

Recently, Dashen, Jenkins, and Manohar suggested an interesting interpretation
of the approximate spin-flavor symmetry of the NRQM ~\cite{dashen,others}.
Working in the large-$N_c$ limit, where $N_c$ is the number of colors, they
showed that the symmetry structure of the baryonic sector of QCD is
constrained by the condition that pion-baryon scattering amplitudes remain
finite as $N_c\rightarrow \infty$, so that unitarity is preserved. Exploiting
these large-$N_c$ consistency conditions, they were able to
classify symmetry-breaking corrections to the mass and decay relations by
their order in the $1/N_c$ expansion.  They observed that the approximate NRQM
spin-flavor structure of the $\ell=0$ baryons in the \su6 \rep{56}
could be understood as a consequence of large-$N_c$, for baryons with small
total spin.  The analogous relations involving baryon states with spins of
order $N_c/2$, however, are subject to large corrections.

Attempts to understand large-$N_c$ baryon phenomenology more directly in terms
of quarks and QCD appeared shortly afterwards in refs.~\cite{carone,luty}.
Ref.~\cite{carone} demonstrates that the connection to quarks
follows from the ideas of Witten (see \cite{witten}), who
showed that large-$N_c$ baryons can be treated in a Hartree
approximation.  In this picture, each quark in the baryon experiences
an average potential generated by the other ${\cal O}(N_c)$ quarks.  In
baryons with small total spin, each quark wavefunction corresponds to
the same s-wave ground state.  In baryons of higher spin, however, the
spin-spin and spin-orbit interactions might significantly deform
the quark wavefunctions away from the s-wave. Ref.~\cite{carone} shows
that this physical picture is consistent with the results of Dashen, Jenkins,
and Manohar.  The Hartree potential, at least in principle, can be computed
using the part of the multiquark Hamiltonian that transforms trivially under
spin and spatial rotations acting separately on each of the quark
wavefunctions.  The remaining piece of the Hamiltonian can then be included
perturbatively.  In this formulation of the problem, the spin-flavor symmetry
appears at lowest order in the $1/N_c$ expansion, and the corrections are
suppressed by powers of $S/N_c$, where $S$ is the baryon spin.  Again, the
approximate spin-flavor symmetry can be understood as a consequence of
large-$N_c$, for baryons with small total spin.

One of the difficulties with the large-$N_c$ picture of baryons is that the
spin and flavor structure of the large-$N_c$ baryons is not simply related to
the spin and flavor structure of the $N_c=3$ baryons, because the
number of quarks is not the same. This has caused considerable confusion
in the literature.  Part of the value of the Hartree picture is that it
suggests a calculational scheme for applying large-$N_c$ ideas to the observed
baryon resonances with $N_c=3$~\cite{carone}.  The first step is to categorize
the relevant multiquark operators by their order in the $1/N_c$ expansion.
This is not completely trivial, since an operator that is summed over the
${\cal O}(N_c)$ quarks in the baryon state may have an effect that is as
important as that of an operator that is formally of lower order, if the terms
in the sum add coherently.  Assuming that we have isolated the correct set of
multiquark operators, we can then apply them to the baryon states,
{\em defined with} $N_c=3$. In this way, we avoid the problem of
extracting our predictions from large-$N_c$ baryon wavefunctions, which
have quantum numbers that are different from those of the baryons in the
real world.

In this paper, we show how to apply these ideas to nonleptonic decays of
the orbitally excited baryons in the \su6 \rep{70}-plet ~\cite{oldpapers}.
In the Hartree language, these are states with $N_c-1$ quarks in the ground
state of the Hartree potential, and one quark in an orbitally excited state.
In contrast with the early work done on this problem, our large-$N_c$
arguments lead us to select a very specific set of pion-baryon interactions.
Furthermore, there is an important difference between these couplings and
those discussed by Dashen, Manohar, and Jenkins for the \rep{56}. In that
case, the leading contribution in large-$N_c$ is identical to the NRQM
prediction (this is related to the fact that the matrix element of the axial
vector current is proportional to $N_c$). However, the dominant decays of the
\rep{70} involve the coupling of pions between the \rep{70} and the \rep{56}.
These matrix elements do not grow with $N_c$. The leading large-$N_c$ result
then includes additional terms beyond those suggested by the NRQM. Thus we can
use our analysis as a test to distinguish between the NRQM and large-$N_c$.
This was one of the motivations of the current work. We hoped to see evidence
that the additional terms included in the large-$N_c$ analysis were necessary
to get an adequate description of the decays. This would have been strong
evidence that large-$N_c$ has something to do with the success of the NRQM.
What we found instead is that the extra terms are not necessary. This result
is inconclusive, in the sense that the coefficients of these terms could be
small even if the large-$N_c$ counting is correct. But the analysis suggests
that there may be more to the NRQM than large-$N_c$.

In the next section, we review the \rep{56} and \rep{70} \su6 representations
of the baryons, as well as their analogues for large $N_c$. We identify the
crucial fact that leads to additional terms in the large-$N_c$ analysis (the
mathematical details are reserved for Appendix~\ref{app:clebsch}). In
Section~\ref{sec:formal}, we discuss our formalism in more detail and present
the set of leading operators.  In Section~\ref{sec:fit} we describe our best
fit to the $\ell=1$ baryon decays.  In Section~\ref{sec:con}, we present our
conclusions.  The technical details of our fits to the known s-wave and d-wave
decay widths are presented in Appendix~\ref{app:fits}.  In
Appendix~\ref{app:pred}, we make predictions for the decay
modes that have not yet been observed and for the modes that have not been
measured precisely.

\section{Preliminaries\label{sec:prelim}}

In this section, we will review the basic elements from \cite{carone} that we
later use to fit the decays of the $\ell=1$ baryons. A more detailed
discussion of these ideas will appear in the next section.

We assume that we can describe the large-$N_c$ baryon states in a tensor
product space of the spin-flavor indices of the $N_c$ valence quarks, as in
the NRQM. Thus our baryons have the spin-flavor and angular momentum structure
of representations of nonrelativistic \su6$\times$O(3). We emphasize that we
are not assuming \su6$\times$O(3).  We are not even trying to make sense of
this as a symmetry group. Rather, we believe that the assumption follows from
a much milder smoothness hypothesis. The argument goes as follows. If the
quarks are very heavy compared to $\Lambda_{QCD}$, the assumption is clearly
correct, because the NRQM description of the baryons can be derived directly
from QCD. The splittings between different spin-flavor states with the same
spatial wavefunctions vanish as the quark masses get large. Thus the states
break up into approximately degenerate multiplets for each spatial
wavefunction. The different spatial wave-functions correspond to different
\su6$\times$O(3) representations. For example, the ground-state wavefunction
is the completely symmetric spin-flavor combination, corresponding to the
Young Tableaux shown in Figure~\ref{youngsym}, with no orbital angular
momentum. The wavefunctions describing the first excited $\ell=1$ baryons
correspond to the Young Tableaux shown in Figure~\ref{youngmix1},
symmetrically
combined with one unit of orbital angular momentum.


The question is, what happens to these approximately degenerate multiplets as
the quarks become light? The thing to notice is that at the bottom of each
multiplet ({\it i.e.} for states with small total spin), the splittings
between
neighboring states are not only suppressed by powers of $1/m_{q}$, but also
by powers of $1/N_c$. Thus, barring some phase transition that leads to a
discontinuous change in the nature of the baryon states, we expect the bottom
of each spin-flavor multiplet to be well-described in the same tensor product
space that works at large $m_q$.  In other words, the NRQM states should be
appropriate.

This argument breaks down at the top of the spin-flavor multiplets, where the
baryon spin is of order $N_c$ and the splittings between neighboring spin
states are of order $\Lambda_{QCD}$ for small quark mass. Thus we expect a
partial spin-flavor symmetry to survive for small quark mass in large-$N_c$.
It is not an approximate symmetry in the usual sense, because symmetry
breaking effects cannot be ignored on any multiplet. Nevertheless, because the
dimensions of the multiplets go to infinity as the small parameter ($1/N_c$)
that characterizes the symmetry breaking goes to zero, we can derive reliable
predictions at one end (for small spin) of the multiplets even though the
symmetry is badly broken at the other. In particular, this argument justifies
the use of the NRQM tensor product states to describe the low-spin baryon
states for large-$N_c$.

While the argument above is theoretically interesting, it leads to one of the
many ambiguities in applying large-$N_c$ arguments to $N_c=3$. How do we
identify states near the ``top'' and ``bottom'' of the multiplets for $N_c=3$?
We will ignore this potential difficulty below and use the expressions we
derive for the entire baryon multiplets. But we should not be surprised if our
results become less reliable as the baryon spin increases.

\subsection*{NRQM versus Large-$N_c$}

Let us now review in more detail the proposal of \cite{carone} for counting
powers of $N_c$. We will do this for matrix elements of operators between
baryon states (the operators could be interpolating fields for mesons),
ignoring flavor symmetry breaking for simplicity. The procedure is simple:
\begin{enumerate}
\item In the spin-flavor space of the NRQM for the baryon states of interest,
write down the most general flavor-conserving expression for the matrix
element.
\item Assign each term in the expression a power of $N_c$ given by the largest
possible power that can appear {\it on the low spin states}. This is most
conveniently determined by simply looking at Feynman diagrams contributing to
the matrix element, making appropriate assumptions about the $N_c$ dependence
of individual quark matrix elements.
\end{enumerate}

Among the Feynman graphs that contribute to the matrix element is a sum over
all quarks of single-quark matrix elements. This has the spin-flavor structure
of the NRQM. In all examples we know of, this gives a contribution to the
leading $N_c$ dependence. The reason that the suggestion above is nontrivial
is that while multiquark diagrams are suppressed by powers of $1/N_c$, their
effects can be enhanced by coherent contributions from the sum over the $N_c$
quarks. This can give additional contributions of the same order in $N_c$ as
the NRQM but with a different spin-flavor structure.

The possible different spin-flavor structures on quark lines can be divided
into four classes:
\begin{enumerate}
\item Constant terms --- these always sum
coherently over the quarks, but the result has no spin-flavor structure
and therefore is not interesting.
\item Spin terms --- these are proportional to
\beq\sum_{\rm quarks\; \it x}\sigma^j_x.\eeq
This never adds coherently on low spin states, so these contributions are
down by $1/N_c$.
\item Flavor terms --- these are proportional to
\beq\sum_{\rm quarks\; \it x}\lambda^a_x.\eeq
This sometimes adds coherently, for example
\beq\sum_{\rm quarks\; \it x}\lambda^8_x\eeq
acting on a low-spin state of $u$ and $d$ quarks is $N_c/\sqrt{12}$.
\item Spin-Flavor terms --- these are proportional to
\beq\sum_{\rm quarks\; \it x}\sigma^j_x\lambda^a_x.\eeq
This can also add coherently; in fact, we show in
Appendix~\ref{app:clebsch} that the \su6 quadratic Casimir operator,
\beq\begin{array}{l}\displaystyle C_2\equiv\left[
{1\over 2f}\sum_{j}\left(\sum_{\rm quarks\; \it
x}\sigma^j_x\right)^2
+{1\over4}\sum_{a}\left(\sum_{\rm quarks\; \it x}\lambda^a_x\right)^2\right.
\\
\displaystyle+\left.{1\over4}\sum_{j,a}\left(\sum_{\rm quarks\; \it
x}\sigma^j_x\lambda^a_x\right)^2
\right]
={2f-1\over 2f}N_c^2(1+{\cal O}(1/N_c))
\end{array}\label{eq:spinflavor}
\eeq
on any finitely excited large-$N_c$ baryon state. Thus, generically, some
spin-flavor matrix elements grow like $N_c$.
\end{enumerate}

As an example of a one-quark contribution, consider the couplings of the
vector mesons, $\rho$ and $\omega$, to the nucleon states. Both couplings grow
with $N_c$, but they are dominated by different contributions. The
contribution to the $\omega$ matrix element is the flavor coupling (in
relativistic notation),
\beq \omega_\mu \,\overline N \,
\left({\textstyle \sum_{x}}\lambda^8_x\right) \,\gamma^\mu N\,,
\label{eq:omega} \eeq
while the leading contribution to the $\rho$ coupling is the spin-flavor
coupling,
\beq (\partial_\mu\rho^a_\nu-\partial_\nu\rho^a_\mu) \,
\overline N \,\left({\textstyle \sum_{x}}
\sigma_x^{\mu\nu}\lambda^a_x\right) \, N\,. \label{eq:rho}
\eeq
The spin-flavor coupling dominates for the $\rho$ coupling because the isospin
matrix element is small for low-spin states, and thus the flavor coupling does
not grow with $N_c$. This is an example of what, in the Skyrme literature, is
called the $I_t=J_t$ rule \cite{mattis,donohue}. Examples in which multiquark
operators contribute at leading order in $N_c$ will appear in the next
section.

\section {Formalism} \label{sec:formal}

We are interested in studying the one-pion decays of \rep{70}-plet baryons
to baryons in the \rep{56}.  While there are also \rep{70} $\rightarrow$
\rep{70} decays, we will not consider them in this paper.  The decays to
states in the \rep{56} are generally favored by the kinematics, and indeed
few \rep{70} $\rightarrow$ \rep{70} modes have been observed in experiment.
In the Hartree language, the part of the interaction Hamiltonian that is of
interest to us can be written
\[
H = \sum _{n=1}^{N_c} \,\,
\sum_{\{x_1 , \ldots , x_n\} \subset \{1, \ldots , N_c\} }
\int d^3 r_{x_1} \ldots d^3 r_{x_n}
\Phi(r_{x_1})_{x_1}^\dagger \otimes \ldots \otimes
\Phi(r_{x_n})_{x_n}^\dagger
\]
\beq
\times {\cal O}(r_{x_1},\ldots ,r_{x_n})
\Psi(r_{x_1})_{x_1} \otimes \ldots \otimes
\Psi(r_{x_{n-1}})_{x_{n-1}} \otimes \Psi_*(r_{x_n})_{x_n}
\label{eq:bigmess}
\eeq
where ${\cal O}$ is the pion coupling to the axial-vector quark current.

Eq. (\ref{eq:bigmess}) requires some explanation.  The $\Phi$s
and $\Psi$s are the individual quark wavefunctions (the self-consistent
solutions to the Hartree equation) for the \rep{56} and \rep{70} baryons,
respectively.  The sum over $n$ indicates that we have broken up the
interaction into parts involving different numbers of quark lines; the
second sum accounts for the possible quark interactions with fixed $n$ that
connect the initial to the final baryon state.  This separation allows
us to classify interactions by their order in the $1/N_c$ expansion.
By Witten's counting arguments, a general $n$-body interaction is of
order $1/N_c^{n-1}$.  For example, there is a distinct term in
(\ref{eq:bigmess}) for the ${\cal O}(1/N_c^2)$ interaction involving three
quark lines shown in Figure~\ref{diagram}.  In the \rep{70} state, one of the
quarks is orbitally excited, which we indicate by the subscript $*$ next to
the ${x_n}^{{\rm th}}$ quark wavefunction.  Notice that each term in
(\ref{eq:bigmess}) involves the wavefunction $\Psi_*$, regardless of the
number of quark lines involved.  This follows because we are only interested
in interactions that contribute to the \rep{70}$\rightarrow$\rep{56} decays,
which necessarily involve the ``de-excitation'' of the orbitally excited
quark.


While it is much too difficult for us to compute the Hartree potential
in a baryon composed of light quarks, we still can learn a great deal
by studying the symmetry structure of (\ref{eq:bigmess}).  As we argued
earlier, it is plausible to represent the small-spin baryon states
made from light quarks in the same space, and by the same representations,
as the baryon states of the naive quark model.  Thus, we work in a
$(2f)^{N_c}$-dimensional tensor product space, where $f$ is the number of
quark flavors.  The quark wavefunctions $\Phi$ and $\Psi$ can be thought of
as $2f \times 2f$ matrices acting on the spin-flavor space of a single
quark; $H$ as a whole can be thought of as a $(2f)^{N_c} \times (2f)^{N_c}$
matrix acting on the $(2f)^{N_c}$-dimensional spin-flavor space in which we
represent the baryon states.  The $\Phi$ and $\Psi$ are the solutions to the
zeroth order Hartree equation, and therefore are spherically symmetric
and spin-flavor independent.

Thus, we can replace these matrix wavefunctions by c-numbers
\beq
\Phi(\vec{r}) \rightarrow \phi(r) \,\,\,\,\, , \,\,\,\,\,
\Psi(\vec{r}) \rightarrow \psi(r) = \phi(r)
\label{eq:cnum}
\eeq
where $r=|\vec{r}|$.  Note that the Hartree potential is a collective
phenomenon and to leading order is unaffected by the excitation of
a single quark.  This accounts for the equality shown in (\ref{eq:cnum}).
In the \rep{70} state, the excited quark has one unit of orbital angular
momentum, so we know the form of its spatial wavefunction:
\beq
\Psi_* (\vec{r}) = f(r) \, Y_{l=1,m}(\theta , \varphi)
= f(r)\,(\vec{r}\cdot \vec{\varepsilon}_m)
\label{eq:epdef}
\eeq
where $f(r)$ is a spin-flavor independent c-number.  In (\ref{eq:epdef})
we have chosen to express the $l=1$ spherical harmonics in terms of
the vectors $\vec{\varepsilon}_m$, which are given by
\beq
\varepsilon_1 = \frac{1}{\sqrt{2}}
\left(\begin{array}{c} -1 \\ -i \\ 0 \end{array} \right)
 \,\,\,
\varepsilon_0 =
\left(\begin{array}{c} 0 \\ 0 \\ 1 \end{array} \right)
\,\,\,
\varepsilon_{-1} = \frac{1}{\sqrt{2}}
\left(\begin{array}{c} 1 \\ -i \\ 0 \end{array} \right)
\eeq
Thus, (\ref{eq:bigmess}) is the integral of the operator
${\cal O}$ times the product of $2N_c$ spherically symmetric functions,
times $\vec{r} \cdot \vec{\varepsilon}_m$.

We can formally perform the integrals once we have specified the
symmetry structure of the operator ${\cal O}$.  In the more familiar
relativistic notation, the pion-quark coupling is given by
\beq
\left(
\overline{q} \gamma^\mu \gamma^5 \lambda^a q
\right)\, \partial_\mu \pi^a / f_{\pi}
\label{eq:avc}
\eeq
where the $\lambda^a$ are \su3 generators.  In the Hartree basis, the
piece of the pion-quark coupling that contributes to baryon decays in
the s-wave has the form
\beq
{\cal O} \sim \lambda^a (\vec{\sigma} \cdot \vec{r}) \, \partial^0
\pi^a / f_{\pi}
\eeq
which, after integration, gives us a one-body interaction that
is leading in $1/N_c$
\beq
a \, \lambda^a_* (\vec{\sigma}_* \cdot \vec{\varepsilon}_m)
\, \partial^0 \pi^a / f_{\pi}
\label{eq:sonebod}
\eeq
where $a$ is an unknown coefficient.  The $*$ under the spin and flavor
matrices indicates that each acts only in the subspace of the orbitally
excited quark.  Recall that a purely one-body interaction must act on the
excited quark line, or there would be no way to change its orbital angular
momentum.  The spin-flavor structure of the operator in (\ref{eq:sonebod})
is consistent with the predictions of the NRQM.

We can also write down a number of operators that are subleading
in $1/N_c$ that involve two quark lines.  However, we will only
include two of these in our subsequent numerical analysis:
\beq
i \, b \, (\vec{\sigma}_* \times \vec{\varepsilon}_m) \cdot
\left(\sum_{x \neq *} \lambda^a_x \vec{\sigma}_x \right)
\, \, \partial^0 \pi^a / f_{\pi}
\label{eq:whynot}
\eeq
\beq
c \, \left(\sum_{x \neq *} \lambda^a_x \right)
(\vec{\sigma}_* \cdot \vec{\varepsilon}_m)
\, \partial^0 \pi^a / f_{\pi}
\label{eq:anotherone}
\eeq
Our motivation for retaining these operators is that the sum over $\lambda^a
\sigma $ in the case of (\ref{eq:whynot}) and the sum over $\lambda^a$
in the case of (\ref{eq:anotherone}) can both be coherent on low-spin
states, and thus the matrix elements can be of
order $1$, rather than order $1/N_c$. This follows from the argument in
Appendix~\ref{app:clebsch}.  Thus we will take our leading s-wave
operators to be those given in (\ref{eq:sonebod}),  (\ref{eq:whynot}),
and (\ref{eq:anotherone}), which we will call operators $A$, $B$, and $C$
respectively.

Arguments analogous to those that we have used to arrive at the operators
responsible for the s-wave decays can also be used to
determine the operators responsible for decays through the d-wave. (Note
that the decay channels in which the pion has odd orbital angular momentum
are forbidden by parity.)  The leading one-body operator is given by
\beq
d \, \lambda^a_* \left( \sigma^{i}_* \varepsilon^{j}_m
+ \sigma^j_* \varepsilon^{i}_m
-\frac{2}{3} \delta^{ij} \vec{\sigma}_* \cdot \vec{\varepsilon}_m \right)
\, \partial^i \partial^j \pi^a / f_{\pi}^2
\label{eq:donebod}
\eeq
We also have two-body operators in the d-wave channel with the same
kind of sum that we encountered in (\ref{eq:whynot})
\beq
i\, e \, \sum_{x\neq *} \left[(\vec{\sigma}_x \times \vec{\sigma}_*)^i
\varepsilon_m^j \lambda^a_x +
(\vec{\sigma}_x \times \vec{\sigma}_*)^j
\varepsilon_m^i \lambda^a_x - \frac{2}{3} (\vec{\sigma}_x \times
\vec{\sigma}_*)
\cdot \vec{\varepsilon}_m \lambda^a_x \delta^{ij} \right]
\, \partial^i \partial^j \pi^a / f_{\pi}^2
\label{eq:dtwobod1}
\eeq
\beq
i \,f \,
\sum_{x \neq *} \left[(\vec{\sigma}_* \times \vec{\varepsilon}_m)^i
\sigma^j_x \lambda^a_x +
(\vec{\sigma}_* \times \vec{\varepsilon}_m)^j
\sigma^i_x \lambda^a_x - \frac{2}{3} (\vec{\sigma}_* \times
\vec{\varepsilon}_m )
\cdot \sigma_x \lambda^a_x \delta^{ij} \right]
\, \partial^i \partial^j \pi^a / f_{\pi}^2
\label{eq:dtwobod2}
\eeq
There is also a third two-body operator involving the cross-product
$(\vec{\sigma}_x \times \vec{\varepsilon}_m)$ which is not linearly
independent of the two operators that we show above.
Finally, there is a d-wave operator analogous to
(\ref{eq:anotherone})
\beq
g \, \left(\sum_{x \neq *} \lambda^a_x \right)
\left( \sigma^{i}_* \varepsilon^{j}_m
+ \sigma^j_* \varepsilon^{i}_m
-\frac{2}{3} \delta^{ij} \vec{\sigma}_* \cdot \vec{\varepsilon}_m \right)
\, \partial^i \partial^j \pi^a / f_{\pi}^2
\label{eq:dtwobod3}
\eeq
Thus, we will retain
(\ref{eq:donebod}), (\ref{eq:dtwobod1}), (\ref{eq:dtwobod2}), and
(\ref{eq:dtwobod3}) as our
set of leading operators in considering the d-wave decays, and refer to
them as operators $D$, $E$, $F$, and $G$.

All that remains is to evaluate our chosen set of operators between the baryon
states, constructed in the $(2f)^{N_c}$-dimensional spin-flavor space.
While the \rep{56} wavefunctions can be represented as completely
symmetric, three-index \su6 tensors, we found it more convenient to use a
six-index notation in which the spin and flavor of each quark are
labeled separately. To represent the \rep{70} states in the most economical
way, we add only two new indices - one which labels the orbital angular
momentum state of the excited quark, and another which tells us which quark
of the three is orbitally excited.  We then check that these
spin-flavor-orbital angular momentum representations of the states are
eigentensors of $J^2$, $J^z$, $I^2$, $I^z$, $\ldots$, with the desired
eigenvalues.  To compute matrix elements, we first act on $n$ quark indices
in the initial baryon state with the desired $n$-body operator, and sum
over the possible combinations; this is equivalent to summing over the
quark lines.  We then compute the inner product of the result with the
tensor representing the final baryon state.  In the next section, we use
matrix elements computed in this way to determine the partial
widths $\Gamma_i ^{({\rm pred})}$, used in our fit of the observed s-wave
and d-wave nonleptonic decays.

\section{Fit\label{sec:fit}}

We must now decide precisely which physical quantities we will fit,
and select the corresponding experimental data.  In addition, we must
arrive at estimates of both the experimental and theoretical uncertainties.
The experimental results we will use are the masses, total decay widths,
and branching fractions given in the 1992 Review of Particle
Properties (RPP) \cite{rpp}. We use the experimentally measured masses,
rather than large-$N_c$ predictions, in computing partial decay widths.  The
masses are affected by large logarithmic corrections proportional to $m_{
\pi}^3/f_\pi^2$ which we would have to include if we were to do the
calculation properly.  For baryons in the \rep{56}, these one-loop corrections
are relatively straightforward to compute, because we know the
mass eigenstates.  For baryons in the \rep{70}, however, we can
determine the mass eigenstates only after including the one-loop corrections.
This makes the problem of computing the masses nonlinear and thus,
far more difficult.  For this reason, the problem of predicting
{\rep{70}}-plet masses in the Hartree picture is best treated separately.

A major problem that we encounter in studying the decay widths
is that the errors in the experimentally determined values of amplitudes at
resonance are often severely underestimated. As a result, one frequently
is presented with two or more mutually inconsistent values for a given decay
channel.  The RPP's approach is to select a few experimental papers that are
considered to be relatively trustworthy, and then to produce an estimated
range of values that is consistent with most or all of these results. A
consequence of this approach is that the uncertainty in the RPP's estimate
of a decay width is generally greater than the error quoted in any of the
experimental papers from which the estimate is derived. It seems to us
that this procedure is safer that the alternative, which is to select one
experimental result for each decay width and then fit our parameters to that
number, ignoring conflicting experimental results. Of course, the large
uncertainties found in the experimental data place a limit on the
precision with which we can extract the underlying parameters.

The values which are generally measured experimentally are the amplitudes
at resonance $\sqrt{\Gamma_i \Gamma_e}/\Gamma_{\tot}$, from which one can
determine the corresponding branching ratios $\Gamma_i / \Gamma_{\tot}$,
provided the elasticity $\Gamma_e / \Gamma_{\tot}$ is known ($\Gamma_e$ is
the partial decay width to the initial state particles used to produce the
resonance).  Unfortunately, the RPP usually provides estimates for the
branching ratios, but not for the amplitudes at resonance. Therefore, it is
the branching ratios which we fit in our analysis. Usually this is not a
problem, as the uncertainty in the elasticity is reasonably small. In a few
cases, however, the elasticity is not very well known, and the uncertainty
propagates to all of the decay fractions for that initial state (the
$\Sigma(1750)$ resonance is an example). For consistency, we do not try to
produce estimates of the amplitude in these situations; instead, we fit the
decay fractions just as we do elsewhere. Finally, we do not attempt to fit
those decay channels for which the RPP does not give an estimate; however,
predictions for these decay modes do appear in Appendix~\ref{app:pred}.

As far as estimates of experimental error are concerned, ranges such as
$10 - 20 \%$ are interpreted as $15 \pm 5 \%$, upper bounds such as
$< 10 \%$ are converted into $5 \pm 5 \%$, and estimates such as
$\approx 0.1 \%$ are interpreted as $0.1 \pm 0.1 \%$. We adopt this scheme
simply as a convention, and not because we believe that any of the
probability distributions are actually gaussian, with the associated
standard deviations. We have found that the precise choice of scheme
for treating the experimental data does not significantly affect our
results.

In addition to fitting the known decay fractions, we simultaneously
fit the total width for each resonance for which at least one decay
channel has been measured. In other words, the quantity we minimize is
\beq \chi^2 = \sum_{{\rm resonance}} \left [{\left
(\Gamma_{\tot}^{(\pred)}-\Gamma_{\tot}^{(\xp)}\right )^2 \over
(\Delta \Gamma_{\tot})^2}+\sum_i {\left ({\Gamma_i^{(\pred)} \over
\Gamma_{\tot}^{(\pred)}}
- f_i^{(\xp)}\right )^2 \over (\Delta f_i)^2}\right ] \eeq
The quantities $\Gamma_{\tot}^{(\pred)}$ are free to vary, whereas the
partial widths $\Gamma_i^{(\pred)}$ are functions of our parameters, namely
the coefficients of the leading $1/N_c$ operators and the mixing angles.
The alternative to this procedure is to hold the total width for each
resonance constant at some best value, and to fit partial widths rather than
decay fractions, combining the uncertainties in the total width and in the
decay fraction to obtain an uncertainty in the partial width. The former
approach is preferred because any uncertainty in a total width
$\Gamma_{\tot}^{(\xp)}$ is only included once, no matter
how many decay channels are measured for that resonance.
As for the data on total widths, we again use the RPP. Just as for the
decay fractions, the RPP's estimates for total widths are quoted as ranges.
Again, we use the midpoint of the range as our best value, and use half the
size of the range as our estimate of the uncertainty.

Another issue to be considered is uncertainty in the masses of some
of the \rep{70}-plet states. For example, the $N(1700)$ mass range is
quoted as 1650 to 1750 MeV. These uncertainties are more important
for d-wave decays than for s-wave decays, because the d-wave kinematic factor
is more sensitive to the initial state mass.  In either case, decays which
occur near threshold are more affected by the precise value of the mass than
those which occur far from threshold.  For the purpose of fitting the data,
we ignore this uncertainty, and simply use the `best' estimate of the mass
quoted in the RPP. However, as we will see in Appendix~\ref{app:pred}, this
possible source of error must be taken into account in our decay predictions.

Theoretical errors also have to be considered. Sources of these errors
include subleading operators in the $1/N_c$ expansion, which we have ignored,
as well as flavor \su3 breaking operators. (The only explicit \su3 breaking
effect that we include is the difference between $f_\pi$ and $f_K$.)  As a
rough estimate, we have assumed a 20\% theoretical uncertainty for each
partial width prediction, and have combined this uncertainty in quadrature
with the experimental uncertainty. The primary effect of this addition is
that the fit is not completely dominated by a few decays which have been
measured extremely well experimentally, in particular, the $\Lambda(1520)$
d-wave decays. For the vast majority of decays, the theoretical error
is not very important, but for consistency we have used the
same value throughout. The choice of a precise value for the theoretical
error does not substantially affect the final results.

\begin{table}[htbp]
\begin{center}
\begin{tabular}{rlccc} \hline\hline
\multicolumn{2}{c}{Decay}&  $f_{{\rm s-wave}}$  & $f_{{\rm d-wave}}$
& $f_ {\rm{exp}}$ \\
\hline\hline
$N(1520)$ & $\rightarrow N \pi $ & - & 65.5 & $55.0 \pm 12.1$ \\
&$\rightarrow N \eta $& - & 0.07 & $0.1 \pm 0.1 $\\
$N(1535)$ & $\rightarrow N \pi $ & 52.6 & -  & $45.0 \pm 13.4$ \\
&$\rightarrow N \eta $& 30.0 & -  & $40.0 \pm 12.8 $\\
&$\rightarrow \Delta \pi $& - & 0.4  & $5.0 \pm 5.0 $\\
$N(1650)$ & $\rightarrow N \pi $ & 78.4 & -  & $70.0 \pm 17.2$ \\
&$\rightarrow N \eta $& 0.9 & -  & $1.0 \pm 1.0 $\\
&$\rightarrow \Lambda K$& 3.2 & - & $7.0 \pm 7.0$ \\
&$\rightarrow \Delta \pi $& - & 9.0  & $5.0 \pm 5.0 $ \\
$N(1675)$ & $\rightarrow N \pi $ & - & 38.3  & $45.0 \pm 10.3$ \\
&$\rightarrow N \eta $& - & 2.1  & $1.0 \pm 1.0 $\\
&$\rightarrow \Lambda K$& - & 0.005 & $0.1 \pm 0.1$ \\
&$\rightarrow \Delta \pi $& - & 53.7  & $55.0 \pm 12.1 $ \\
$N(1700)$ & $\rightarrow N \pi $ & - & 13.2  & $10.0 \pm 5.4$ \\
&$\rightarrow \Lambda K$& - & 0.09 & $0.2 \pm 0.1$ \\
$\Delta(1620)$ & $\rightarrow N \pi $ & 18.7 & -  & $25.0 \pm 7.1$ \\
&$\rightarrow \Delta \pi $& - & 41.8  & $50.0 \pm 14.1 $ \\
$\Delta(1700)$ & $\rightarrow N \pi $ & - & 12.0  & $15.0 \pm 5.8$ \\
\hline \hline
\end{tabular}
\caption{Predicted branching fractions, corresponding to the parameter
set $a= 0.536$, $b=-0.028$, $c=0.101$, $d=0.203$, $e=-0.015$, $f=-0.029$,
$g=-0.002$, and the mixing angles $\theta_{N1}=0.61$, $\theta_{N3}=3.04$,
$\theta_{\Lambda11}=1.78$, $\theta_{\Lambda12}=2.79$,
$\theta_{\Lambda13}=1.53$, $\theta_{\Lambda31}=0.32$,
$\theta_{\Lambda32}=0.14$,
$\theta_{\Lambda33}=2.63$, $\theta_{\Sigma11}=2.00$,
$\theta_{\Sigma12}=1.16$,
$\theta_{\Sigma31}=2.14$, $\theta_{\Sigma32}=0.48$\label{t1}}
\end{center}
\end{table}

\begin{table}[htbp]
\begin{center}
\begin{tabular}{rlccc} \hline \hline
\multicolumn{2}{c}{Decay}&  $f_{{\rm s-wave}}$  & $f_{{\rm d-wave}}$
& $f_ {\rm{exp}}$ \\
\hline\hline
$\Lambda(1520)$ & $\rightarrow N \overline{K} $ & - & 17.9
& $45.0 \pm 9.1$ \\
&$\rightarrow \Sigma \pi  $& - & 41.5  & $42.0 \pm 8.5 $\\
$\Lambda(1670)$ & $\rightarrow N \overline{K} $ & 20.2 & -
& $20.0 \pm 6.4$ \\
&$\rightarrow \Sigma \pi  $& 40.2 & -  & $40.0 \pm 21.5 $\\
&$\rightarrow \Lambda \eta$& 25.1 &-& $25.0 \pm 11.2$ \\
$\Lambda(1690)$ & $\rightarrow N \overline{K} $ & - & 21.7
& $25.0 \pm 7.1$ \\
&$\rightarrow \Sigma \pi  $& - & 30.3  & $30.0 \pm 11.7 $\\
$\Lambda(1800)$ & $\rightarrow N \overline{K} $ & 32.6 & -
& $32.5 \pm 9.9$ \\
$\Lambda(1830)$ & $\rightarrow N \overline{K} $ & - & 1.3
& $6.5 \pm 3.7$ \\
&$\rightarrow \Sigma \pi  $& - & 83.2  & $55.0 \pm 22.8 $\\
$\Sigma(1670)$ & $\rightarrow N \overline{K} $ & - & 4.0
& $10.0 \pm 3.6$ \\
&$\rightarrow \Lambda \pi  $& - & 11.6  & $10.0 \pm 6.4 $\\
&$\rightarrow \Sigma \pi  $& - & 44.4  & $45.0 \pm 17.5 $\\
$\Sigma(1750)$ & $\rightarrow N \overline{K} $ & 28.1 & -
& $25.0 \pm 15.8$ \\
&$\rightarrow \Sigma \pi  $& 4.2& -  & $4.0 \pm 4.0 $\\
&$\rightarrow \Sigma \eta  $& 6.5 & -  & $35.0 \pm 21.2 $\\
$\Sigma(1775)$ & $\rightarrow N \overline{K} $ & - & 17.3
& $40.0 \pm 8.5$ \\
&$\rightarrow \Lambda \pi  $& - & 25.6  & $17.0 \pm 4.5 $\\
&$\rightarrow \Sigma \pi  $& - & 3.4  & $3.5 \pm 1.7 $\\
&$\rightarrow \Sigma^* \pi  $& - & 6.7  & $10.0 \pm 2.8 $\\
\hline \hline
\end{tabular}
\centerline{Table 1: (continued)}
\end{center}
\end{table}

Note that the estimates for the different decay fractions of a given resonance
are not really independent of each other, even though we treat them as such
for the purpose of the fit. At the end, we must check that our predicted
values for both measured and unmeasured decay widths, together with
measured non \rep{56}-pion decay widths, add up to the
full width to within the allowed uncertainties. In cases where the
non \rep{56}-pion decays are poorly known, we must at least ascertain
that the predicted \rep{56}-pion decay fractions sum to a number less
than unity. Further details are discussed in Appendices~\ref{app:fits} and
\ref{app:pred}.

In Table~\ref{t1} we show the best fit for the measured decays
that go entirely through one partial wave. Other fits, involving
different mixing angles but very similar values of the parameters
$a$, $b$, $c$, $d$, $e$, $f$, and $g$,
are discussed in Appendix~\ref{app:fits}.
The definitions of the mixing angles also appear in Appendix~\ref{app:fits}.
The quality of the fits is reasonable (the pure s-wave fit has
a $\chi^2 = 4.5$ for $4$ degrees of freedom, while the pure
d-wave fit has a $\chi^2 = 36.0$ for $15$ degrees of freedom). With
a few exceptions (notable ones being the $\Lambda(1520) \rightarrow
N \overline{K}$ and $\Sigma(1775) \rightarrow N \overline{K}$ decays),
the predictions are within the range of uncertainty given by the combined
experimental and theoretical errors. The most interesting feature of the fit
presented in Table~\ref{t1} is the smallness of parameters $b$ and $c$
relative to $a$ and of parameters $e$, $f$, and $g$ relative to $d$.
This will be discussed further in the following section.

\section{Conclusions} \label{sec:con}

We have shown how to compute the leading one-pion decay amplitudes
for the orbitally excited, \rep{70}-plet baryons in the
large-$N_c$ limit.  By working in a Hartree approximation, we arrived
a specific set of operators that are responsible for decays through the
s-wave and d-wave channels.  While the fits we obtained to the current
experimental data were not necessarily better than those obtained by others
using different methods, our results have the advantage of following more
directly from the underlying physics in a well-defined limit of QCD.

A striking feature of our results is the suppression of the
two-body operators, $B$, $C$, $E$, $F$, and $G$.  Since these operators are
one higher order in the $1/N_c$ expansion than $A$ and $D$, we
expected a relative suppression in their coefficients, compensated by an
enhancement in the matrix elements.  The interesting
point is that this suppression was generally much greater than
a factor of $N_c=3$.  The two-body operators that we retained all
involved a sum over quark lines which we argued should lead to
an enhancement of order $N_c$.  However, the values of
$b$, $c$, $e$, $f$, and $g$ that we obtained in the fits were so small that
the matrix elements of the two-body operators are suppressed
even when the sums over quark lines are coherent.

One possible conclusion from this result is that there is something more to
the success of the NRQM for baryons than large-$N_c$. Perhaps somehow, in
spite of the fact that the quarks are not really heavy, they act in the
process of $\ell=1$ baryon decay as if they were.

\centerline{{\bf Acknowledgments}}

We thank Aneesh Manohar, Michael Mattis and Sam Osofsky for useful
conversations. We are also grateful to Sam Osofsky for double-checking our
baryon wavefunctions.  {\it  This research was supported in part by the
National Science Foundation, under grant PHY-9218167, and in part by the Texas
National Research Laboratory Commission, under grant RGFY93-278B.}


\appendix
\section{$N_c$ Dependence of Spin-Flavor Generators\label{app:clebsch}}

In this section we derive (\ref{eq:spinflavor}). The Casimir operator can be
written
\beq C_2\equiv\sum_\alpha\,T_\alpha^2 \label{eq:a1}\eeq
where the $T_\alpha$ are the \su{2f} generators, normalized so that
\beq {\rm tr\,} T_\alpha T_\beta=\delta_{\alpha\beta}\label{eq:a2}\eeq
in the defining, $2f$ dimensional representation. Rather than computing the
Casimir operator directly in other representations, $R$, it is easier to
compute the quantity $T(R)$, defined by
\beq {\rm tr}_R\,T_\alpha T_\beta=T(R)\,\delta_{\alpha\beta}\label{eq:a3}.
\eeq
Then $C_2$ can be obtained as follows:
\beq C_2=(4f^2-1){T(R)\over D(R)}\,,\label{eq:a4}\eeq
where $D(R)$ is the dimension of the representation, $R$. Thus, for example,
in the defining representation, the Casimir operator is
\beq C_2={4f^2-1\over2f}\,.\label{eq:a5}\eeq

The crucial step in obtaining (\ref{eq:spinflavor}) is to calculate $T(R)$ for
the completely symmetric representation of Figure~\ref{youngsym}. Let us call
this representation $\{N_c\}$. We will calculate the trace of the square of a
generator that is the analogue of $\lambda^8$ for \su{2f},
\beq T_{2f-1}\equiv{1\over\sqrt{2f(2f-1)}}\pmatrix{
1&0&\cdots&0\cr
0&1&\cdots&0\cr
\vdots&\vdots&\ddots&\vdots\cr
0&0&\cdots&1-2f\cr}\label{eq:a6}\eeq
Then we can compute the trace by noting that in $\{N_c\}$, there are
\beq
\left({N_c+2f-k-2\atop N_c-k}\right)
\label{eq:a7}\eeq
states with $k$ indices having value $2f$, on each of which the value of
$T_{2f-1}^2$
is
$${1 \over 2f(2f-1)}[k(1-2f)+(N_c-k)]^2,$$ thus
$$T(\{N_c\})={\rm tr}_{\{N_c\}} T_{2f-1}^2$$
\beq
={1\over2f(2f-1)}\sum_{k=0}^{N_c} [k(2f-1)-(N_c-k)]^2
\left({N_c+2f-k-2\atop N_c-k}\right)
\label{eq:a8}\eeq
$$=\left({N_c+2f\atop N_c-1}\right)$$
This gives
\beq
C_2(\{N_c\})={2f-1\over2f}\,N_c(N_c+2f)
\label{eq:a9}\eeq
in agreement with (\ref{eq:spinflavor}).

The reason that (\ref{eq:spinflavor}) is correct for any finitely excited
baryon state is that the order $N_c^2$ term comes from the horizontal string
of boxes in the Young tableaux with length of order $N_c$, a feature shared by
all the finitely excited large-$N_c$ baryon states. More precisely, note that
(\ref{eq:a9}) implies
\beq
C_2(\{N_c-\ell\})={2f-1\over2f}\,N_c^2+{\cal O}(N_c)
\label{eq:a10}\eeq
for any fixed $\ell$ as $N_c\rightarrow\infty$. Note further that we can
determine $T(R)$ for any finitely excited baryon state by starting with the
representations, $\{N_c-\ell\}$, and using the recursion relations
\beq
T(r\otimes R)=D(r)T(R)+D(R)T(r)\,,\quad
T(r\oplus R)=T(r)+T(R)\,.
\label{eq:a11}\eeq
The point is that the Clebsch-Gordon decomposition in (\ref{eq:a11}) does not
change $C_2$ to leading order in $N_c$ because $T(\{N_c-\ell\})$ is higher
order in $N_c$ than $D(\{N_c-\ell\})$. Thus
$${T(r\otimes \{N_c-\ell\})\over
D(r\otimes \{N_c-\ell\})}
={D(r)T(\{N_c-\ell\})+D(\{N_c-\ell\})T(r)\over D(r)D(\{N_c-\ell\})}$$
\beq
={T(\{N_c-\ell\})\over D(\{N_c-\ell\})}+{\cal O}(1)
\label{eq:a12}\eeq
for any fixed $r$. Then the standard rules of Clebsch-Gordon decomposition can
be used to establish (\ref{eq:spinflavor}) for any representation obtained
from $\{N_c-\ell\}$ by adding a finite number of boxes.

\section{Fits of known decay data} \label{app:fits}
\centerline{{\em S-wave decays}}

We first consider decay channels that are pure s-wave, that is,
where both the \rep{70} and \rep{56} baryon states have spin $1/2$.
Thirteen such decays have been measured, associated with six
\rep{70}-plet resonances. The data is presented in
Tables~\ref{snomix}-\ref{ssigma}. The
$\Lambda(1405) \rightarrow \Sigma \pi$ decay has been omitted from the fit
because it is questionable whether the $\Lambda(1405)$ can be described
in the \su6 model. In particular, one suspects that the $\Lambda(1405)$
may consist largely of an unstable $N\overline{K}$ bound
state.~\cite{molecule}
Our
prediction for the $\Lambda(1405) \rightarrow \Sigma \pi$ decay rate,
$ 0 - 10 $ MeV, based on the assumption that the $\Lambda(1405)$ is
an SU(6) state orthogonal to $\Lambda(1670)$ and $\Lambda(1800)$,
is in fact much smaller than the measured value, $50$ MeV.

\begin{table}[htbp]
\begin{center}
\begin{tabular}{lcc} \hline \hline
& $\Gamma_{\tot}^{(\xp)}(\MeV)$ & $\Gamma_{\tot}^{(\pred)}(\MeV)$ \\
$\Delta(1620)$ & $150 \pm 30$ & $134.2$ \\ \hline
& $f_i^{(\xp)}(\%)$ & $f_i^{(\pred)}(\%)$ \\
$\rightarrow N \pi$ & $25.0 \pm 7.1$ & $18.7$ \\ \hline \hline
& $\Gamma_{\tot}^{(\xp)}(\MeV)$ & $\Gamma_{\tot}^{(\pred)}(\MeV)$ \\
$N(1535)$ & $175 \pm 75$ & $186.5$ \\ \hline
& $f_i^{(\xp)}(\%)$ & $f_i^{(\pred)}(\%)$ \\
$\rightarrow N \pi$ & $45.0 \pm 13.4$ & $52.6$ \\
$\rightarrow N \eta$ & $40.0 \pm 12.8$ & $30.0$ \\ \hline \hline
& $\Gamma_{\tot}^{(\xp)}(\MeV)$ & $\Gamma_{\tot}^{(\pred)}(\MeV)$ \\
$N(1650)$ & $167.5 \pm 22.5$ & $173.1$ \\ \hline
& $f_i^{(\xp)}(\%)$ & $f_i^{(\pred)}(\%)$ \\
$\rightarrow N \pi$ & $70.0 \pm 17.2$ & $78.4$ \\
$\rightarrow N \eta$ & $1.0 \pm 1.0$ & $0.9$ \\
$\rightarrow \Lambda K$ & $7.0 \pm 7.1$ & $3.2$ \\ \hline \hline
\end{tabular}
\caption{S-wave decays for $\Delta$ and $N$ initial states}
\label{snomix}
\end{center}
\end{table}

\begin{table}[htbp]
\begin{center}
\begin{tabular}{lccccc} \hline \hline
& & fits \#1,6 & fits \#2,4 & fits \#3,5 & fits \#7,8 \\ \hline \hline
& $\Gamma_{\tot}^{(\xp)}(\MeV)$ &
\multicolumn{4}{c}{$\Gamma_{\tot}^{(\pred)}(\MeV)$} \\
$\Lambda(1670)$ & $37.5 \pm 12.5$ & 38.1 & 38.0 & 32.3 & 31.5 \\ \hline
& $f_i^{(\xp)}(\%)$ & \multicolumn{4}{c}{$f_i^{(\pred)}(\%)$} \\
$\rightarrow N\overline{K}$ & $20.0 \pm 6.4$ & 20.2 & 19.9 & 20.1 & 19.1 \\
$\rightarrow \Sigma \pi$ & $40.0 \pm 21.5$ & 40.2 & 39.3 & 37.1 & 40.1 \\
$\rightarrow \Lambda \eta$ & $25.0 \pm 11.2$ & 25.1 & 26.1 & 19.2 & 20.1 \\
\hline \hline
& $\Gamma_{\tot}^{(\xp)}(\MeV)$ &
\multicolumn{4}{c}{$\Gamma_{\tot}^{(\pred)}(\MeV)$} \\
$\Lambda(1800)$ & $300 \pm 100$ & 300.7 & 300.6 & 299.0 & 299.7 \\ \hline
& $f_i^{(\xp)}(\%)$ & \multicolumn{4}{c}{$f_i^{(\pred)}(\%)$} \\
$\rightarrow N \overline{K}$ & $32.5 \pm 9.9$ & 32.6 & 32.6 & 32.4 & 32.5 \\
\hline \hline
\end{tabular}
\caption{S-wave decays for $\Lambda$ initial states}
\label{slambda}
\end{center}
\end{table}

\begin{table}[htbp]
\begin{center}
\begin{tabular}{lccc} \hline \hline
& & fit \#1,2 & fit \#3,4 \\ \hline \hline
& $\Gamma_{\tot}^{(\xp)}(\MeV)$ &
\multicolumn{2}{c}{$\Gamma_{\tot}^{(\pred)}(\MeV)$} \\
$\Sigma(1750)$ & $110 \pm 50$ & $109.7$ & $110.1$ \\ \hline
& $f_i^{(\xp)}(\%)$ & \multicolumn{2}{c}{$f_i^{(\pred)}(\%)$} \\
$\rightarrow N\overline{K}$ & $25.0 \pm 15.8$ & 28.1 & 27.3 \\
$\rightarrow \Sigma \pi$ & $4.0 \pm 4.1$ & 4.2 & 3.8 \\
$\rightarrow \Sigma \eta$ & $35.0 \pm 21.2$ & 6.5 & 2.7 \\ \hline \hline
\end{tabular}
\caption{S-wave decays for $\Sigma$ initial states}
\label{ssigma}
\end{center}
\end{table}

Our conventions for the mixing angles are as follows.  One angle
($\theta_{N1}$) is needed to specify the spin-$1/2$ nucleon states:
\beq \left [\begin {array}{c} {N(1535)}\\\noalign{\medskip}{N(1650)}\end
{array}\right ] =
\left [\begin {array}{cc} {\cos(\theta_{N1})}&{\sin(\theta_{N1})}
\\\noalign{\medskip}{-\sin(\theta_{N1})}&{\cos(\theta_{N1})}\end {array}
\right ]
\left [\begin {array}{c} {N_{11}}\\\noalign{\medskip}{N_{31}}
\end {array}\right ] \eeq
where our convention for the pure \su6 states on the right hand side
is that the first subscript is twice the total quark spin
of the baryon state, and the second is twice the total angular momentum.
Three angles ($\theta_{\Lambda 1i}$, $i=1..3$) are used for the $\Lambda$
mixing matrix:
\[ \left [\begin {array}{c} {\Lambda(1670)}\\\noalign{\medskip}
{\Lambda(1800)}\\\noalign{\medskip}{\Lambda(1405)}
\end {array}\right ] = \]
\beq \eeq
\[
\left [\begin {array}{ccc}
{c_{\Lambda 11}c_{\Lambda 12}}&
{s_{\Lambda 11}c_{\Lambda 12}}&
{s_{\Lambda 12}}\\\noalign{\medskip}
{-s_{\Lambda 11}c_{\Lambda 13}
-c_{\Lambda 11}s_{\Lambda 13}s_{\Lambda 12}}&
{c_{\Lambda 11}c_{\Lambda 13}
-s_{\Lambda 11}s_{\Lambda 13}s_{\Lambda 12}}&
{s_{\Lambda 13}c_{\Lambda 12}}
\\\noalign{\medskip}
{s_{\Lambda 11}s_{\Lambda 13}
-c_{\Lambda 11}c_{\Lambda 13}s_{\Lambda 12}}&
{-c_{\Lambda 11}s_{\Lambda 13}
-s_{\Lambda 11}c_{\Lambda 13}s_{\Lambda 12}}&
{c_{\Lambda 13}c_{\Lambda 12}}\end {array}\right ]
\left [\begin {array}{c} {\Lambda_{11}}\\\noalign{\medskip}{\Lambda_{31}}
\\\noalign{\medskip}{{\rm Singlet}_{11}}
\end {array}\right ] \]
where we use the abbreviation $c_{\Lambda 11} = \cos(\theta_{\Lambda 11})$,
etc.  Finally, because we have decay data for only one of the three physical
spin-$1/2$ $\Sigma$ states, only two mixing angles
($\theta_{\Sigma 11}$ and $\theta_{\Sigma 12}$) are needed:
\beq \Sigma(1750)= \left [\begin {array}{ccc}
c_{\Sigma 11}c_{\Sigma 12} &
s_{\Sigma 11}c_{\Sigma 12} &
s_{\Sigma 12} \end {array} \right ]
\left [\begin {array}{c} {\Sigma_{11}}\\\noalign{\medskip}{\Sigma_{31}}
\\\noalign{\medskip}{\Sigma^{\ast}_{11}}
\end {array}\right ] \eeq
Our conventions for all of the mixing matrices and angles are such that if
the RPP assignments of the \rep{70}-plet states in the quark model were
correct, all the mixing matrices would be diagonal and all of the angles would
equal 0 (in our fit, we have in fact chosen all of the angles to
lie in the interval $[0,\pi)$).

As discussed in Section~\ref{sec:formal}, we expect three operators
$A$, $B$, and $C$
to contribute to s-wave decays at leading order in the $1/N_c$ expansion.
Thus, we must fit a total of nine parameters (the three coefficients $a$,
$b$, and $c$ in addition to the six mixing angles) to thirteen decay
fractions, leaving us with four degrees of freedom. The best fit produces
a $\chi^2$ of $4.47$.
However, there are a number of minima with $\chi^2$ close to its minimum
value, which all have roughly the same values for the parameters
$a$, $b$, and $c$,
but have different values for the various mixing angles. We found only one
solution for the nucleon mixing angle $\theta_{N1}$, eight possible
solutions for the $\Lambda$ mixing matrix, and four solutions for
the $\Sigma$ mixing matrix. All of the solutions are tabulated in Table~
\ref{spars}. The quantity $\Delta \chi^2$ associated with each solution
is computed relative to our best solution, which has
$\chi^2=4.47$. For example, if we choose fit \#3 for the $\Lambda$ angles and
fit \#4 for the $\Sigma$ angles, we obtain a total $\chi^2$ of $5.22$.
Table~\ref{spars} also lists the uncertainties in all of the parameters, as
obtained from the covariance matrix.  The calculated values of the decay
fractions corresponding to each of the solutions are listed in
Tables~\ref{snomix}-\ref{ssigma} for comparison with experimental data.  We
present in Tables~\ref{snucmix}-\ref{ssigmix} the spin-$1/2$ $N$, $\Lambda$,
and $\Sigma$ mixing matrices corresponding to the various solutions, along
with associated uncertainties.


\begin{table}[htbp]
\begin{center}
\begin{tabular}{lcccc} \hline \hline
Parameter & value \\ \hline \hline
a & $0.536 \pm 0.071$ \\
b & $-0.028 \pm 0.022$ \\
c & $0.101 \pm 0.059$ \\
$\theta_{N1}$ & $0.61 \pm 0.09$ \\ \hline \hline
& fit \#1 & fit \#2 & fit \#3 & fit \#4 \\
$\theta_{\Lambda 11}$ & $1.78 \pm 0.15$ & $1.33 \pm 0.19$ & $0.99 \pm 0.18$ &
$1.33 \pm 0.19$ \\
$\theta_{\Lambda 12}$ & $2.79 \pm 0.10$ & $2.19 \pm 0.15$ & $2.41 \pm 0.15$ &
$2.19 \pm 0.15$ \\
$\theta_{\Lambda 13}$ & $1.53 \pm 0.20$ & $1.70 \pm 0.19$ & $1.96 \pm 0.20$ &
$2.71 \pm 0.22$ \\
$\Delta \chi^2$ & 0.00 & 0.01 & 0.25 & 0.01 \\ \hline
& fit \#5 & fit \#6 & fit \#7 & fit \#8 \\
$\theta_{\Lambda 11}$ & $0.99 \pm 0.18$ & $1.78 \pm 0.15$ & $1.46 \pm 0.15$ &
$1.46 \pm 0.15$ \\
$\theta_{\Lambda 12}$ & $2.41 \pm 0.15$ & $2.79 \pm 0.10$ & $2.88 \pm 0.10$ &
$2.88 \pm 0.10$ \\
$\theta_{\Lambda 13}$ & $2.97 \pm 0.21$ & $2.54 \pm 0.22$ & $2.64 \pm 0.21$ &
$1.63 \pm 0.20$ \\
$\Delta \chi^2$ & 0.25 & 0.00 & 0.27 & 0.27 \\ \hline \hline
& fit \#1 & fit \#2 & fit \#3 & fit \#4 \\
$\theta_{\Sigma 11}$ & $2.00 \pm 0.29$ & $2.18 \pm 0.10$ & $0.77 \pm 0.91$ &
$1.97 \pm 0.14$ \\
$\theta_{\Sigma 12}$ & $1.16 \pm 0.47$ & $3.01 \pm 0.47$ & $1.29 \pm 0.12$ &
$2.65 \pm 0.31$ \\
$\Delta \chi^2$ & 0.00 & 0.00 & 0.50 & 0.50 \\ \hline \hline
\end{tabular}
\caption{Parameters from s-wave fit}
\label{spars}
\end{center}
\end{table}

\begin{table}[htbp]
\[ M_{N1} = \left [\begin {array}{cc}
0.82 \pm 0.05 & 0.57 \pm 0.07
\\\noalign{\medskip}
-0.57 \pm 0.07 & 0.82 \pm 0.05
\end {array}
\right ] \]
\caption{Spin-$1/2$ nucleon mixing matrix}
\label{snucmix}
\end{table}

\begin{table}[htbp]
\[ M_{\Lambda 1}^{(\fit \#1)} = \left [\begin {array}{ccc}
0.19 \pm 0.14 & -0.92 \pm 0.04 & 0.34 \pm 0.09
\\\noalign{\medskip}
0.03 \pm 0.19 & -0.35 \pm 0.11 & -0.94 \pm 0.04
\\\noalign{\medskip}
0.98 \pm 0.03 & 0.19 \pm 0.16 & -0.04 \pm 0.19
\end {array}\right ] \]
\[ M_{\Lambda 1}^{(\fit \#2)} = \left [\begin {array}{ccc}
-0.14 \pm 0.12 & -0.56 \pm 0.11 & 0.81 \pm 0.09
\\\noalign{\medskip}
-0.07 \pm 0.20 & -0.81 \pm 0.08 & -0.58 \pm 0.12
\\\noalign{\medskip}
0.99 \pm 0.03 & -0.13 \pm 0.20 & 0.07 \pm 0.12
\end {array}\right ] \]
\[ M_{\Lambda 1}^{(\fit \#3)} = \left [\begin {array}{ccc}
-0.41 \pm 0.14 & -0.62 \pm 0.09 & 0.67 \pm 0.11
\\\noalign{\medskip}
-0.02 \pm 0.18 & -0.72 \pm 0.10 & -0.69 \pm 0.11
\\\noalign{\medskip}
0.91 \pm 0.06 & -0.30 \pm 0.19 & 0.28 \pm 0.14
\end {array}\right ] \]
\[ M_{\Lambda 1}^{(\fit \#4)} = \left [\begin {array}{ccc}
-0.14 \pm 0.12 & -0.56 \pm 0.11 & 0.81 \pm 0.09
\\\noalign{\medskip}
0.80 \pm 0.09 & -0.55 \pm 0.12 & -0.24 \pm 0.11
\\\noalign{\medskip}
0.58 \pm 0.13 & 0.62 \pm 0.10 & 0.53 \pm 0.14
\end {array}\right ] \] \\
\centerline{Table \ref{slammix}: Spin-$1/2$ $\Lambda$ mixing matrices}
\end{table}

\begin{table}[htbp]
\[ M_{\Lambda 1}^{(\fit \#5)} = \left [\begin {array}{ccc}
-0.41 \pm 0.14 & -0.62 \pm 0.09 & 0.67 \pm 0.11
\\\noalign{\medskip}
0.76 \pm 0.08 & -0.64 \pm 0.09 & -0.13 \pm 0.15
\\\noalign{\medskip}
0.50 \pm 0.14 & 0.46 \pm 0.12 & 0.73 \pm 0.11
\end {array}\right ] \]
\[ M_{\Lambda 1}^{(\fit \#6)} = \left [\begin {array}{ccc}
0.19 \pm 0.14 & -0.92 \pm 0.04 & 0.34 \pm 0.09
\\\noalign{\medskip}
0.85 \pm 0.11 & -0.02 \pm 0.14 & -0.53 \pm 0.17
\\\noalign{\medskip}
0.49 \pm 0.16 & 0.40 \pm 0.09 & 0.77 \pm 0.13
\end {array}\right ] \]
\[ M_{\Lambda 1}^{(\fit \#7)} = \left [\begin {array}{ccc}
-0.11 \pm 0.14 & -0.96 \pm 0.03 & 0.26 \pm 0.10
\\\noalign{\medskip}
0.86 \pm 0.09 & -0.22 \pm 0.11 & -0.46 \pm 0.18
\\\noalign{\medskip}
0.50 \pm 0.17 & 0.17 \pm 0.10 & 0.85 \pm 0.11
\end {array}\right ] \]
\[ M_{\Lambda 1}^{(\fit \#8)} = \left [\begin {array}{ccc}
-0.11 \pm 0.14 & -0.96 \pm 0.03 & 0.26 \pm 0.10
\\\noalign{\medskip}
0.03 \pm 0.19 & -0.26 \pm 0.10 & -0.96 \pm 0.03
\\\noalign{\medskip}
1.00 \pm 0.02 & -0.10 \pm 0.14 & 0.06 \pm 0.19
\end {array}\right ] \]
\caption{Spin-$1/2$ $\Lambda$ mixing matrices (continued)}
\label{slammix}
\end{table}

\begin{table}[htbp]
\[ M_{\Sigma 1}^{(\fit \#1)} = \left [\begin {array}{ccc}
-0.17 \pm 0.27 & 0.36 \pm 0.35 & 0.92 \pm 0.19
\end {array}\right ] \]
\[ M_{\Sigma 1}^{(\fit \#2)} = \left [\begin {array}{ccc}
0.57 \pm 0.07 & -0.81 \pm 0.09 & 0.13 \pm 0.46
\end {array}\right ] \]
\[ M_{\Sigma 1}^{(\fit \#3)} = \left [\begin {array}{ccc}
0.20 \pm 0.15 & 0.19 \pm 0.23 & 0.96 \pm 0.03
\end {array}\right ] \]
\[ M_{\Sigma 1}^{(\fit \#4)} = \left [\begin {array}{ccc}
0.34 \pm 0.11 & -0.81 \pm 0.16 & 0.47 \pm 0.27
\end {array}\right ] \]
\caption{Spin-$1/2$ $\Sigma$ mixing matrices}
\label{ssigmix}
\end{table}

We note from Table~\ref{spars} that the coefficient $b$ is
strongly suppressed relative to $a$, more than one might expect from naive
$1/N_c$ power counting, with $N_c=3$.
Because of the uncertainty in the value of $c$ obtained from the fit,
it is not as clear that $c$ is strongly suppressed. However, consideration
of s+d-wave decays in Appendix~\ref{app:pred} leads to to believe that
$c$ is in fact near the lower end of the range presented in
Table~\ref{spars}.
For the fitted value of $\theta_{N1}$,
we see that there is significant mixing between the $N_{11}$ and $N_{31}$
states. It is somewhat difficult to draw conclusions about the mixing of the
$\Lambda$ and $\Sigma$ resonances, due to the presence of multiple solutions.
For example, fits \#1 and \#8 for the $\Lambda$ angles predict little mixing,
but with assignments for the three states different from those given in the
RPP. Fits \#6 and \#7 also predict a limited amount of mixing, but with the
identification of $\Lambda(1800)$ and $\Lambda(1405)$ reversed. Fits \#3-5 all
predict a substantial amount of mixing. As far as the $\Sigma$ states, it is
not possible to say definitively whether the $\Sigma(1750)$ consists mostly of
$\Sigma_{31}$ or of $\Sigma_{11}^{\ast}$.

In Table~\ref{slambda}, we see that the most obvious difference between fits
\#1,2,4,6 and fits \#3,5,7,8 is that the latter predict a smaller partial
width for $\Lambda(1670) \rightarrow \Lambda \eta$. In Table~\ref{ssigma}, we
notice that the main problem with $\Sigma(1750)$ decays is to obtain a
reasonable value for the $\Sigma \eta$ channel.

\centerline{{\em D-wave decays}}

The same procedure is followed as for the s-wave decay rates. Here, we
only fit those decays which are pure d-wave, that is, we omit spin-$3/2$
to spin-$3/2$ decays (these will be discussed in Appendix~\ref{app:pred}).
Our conventions for the mixing angles of the spin-$3/2$ states are analogous
to those for the spin-$1/2$ states. For the nucleons,
\beq \left [\begin {array}{c} {N(1520)}\\\noalign{\medskip}{N(1700)}\end
{array}\right ] =
\left [\begin {array}{cc} {\cos(\theta_{N3})}&{\sin(\theta_{N3})}
\\\noalign{\medskip}{-\sin(\theta_{N3})}&{\cos(\theta_{N3})}\end {array}
\right ]
\left [\begin {array}{c} {N_{13}}\\\noalign{\medskip}{N_{33}}
\end {array}\right ] \eeq
For the $\Lambda$ states,
\[ \left [\begin {array}{c} {\Lambda(1690)}\\\noalign{\medskip}
{\Lambda(??)}\\\noalign{\medskip}{\Lambda(1520)}
\end {array}\right ] = \]
\beq\eeq
\[
\left [\begin {array}{ccc}
{c_{\Lambda 31}c_{\Lambda 32}}&
{s_{\Lambda 31}c_{\Lambda 32}}&
{s_{\Lambda 32}}\\\noalign{\medskip}
{-s_{\Lambda 31}c_{\Lambda 33}
-c_{\Lambda 31}s_{\Lambda 33}s_{\Lambda 32}}&
{c_{\Lambda 31}c_{\Lambda 33}
-s_{\Lambda 31}s_{\Lambda 33}s_{\Lambda 32}}&
{s_{\Lambda 33}c_{\Lambda 32}}
\\\noalign{\medskip}
{s_{\Lambda 31}s_{\Lambda 33}
-c_{\Lambda 31}c_{\Lambda 33}s_{\Lambda 32}}&
{-c_{\Lambda 31}s_{\Lambda 33}
-s_{\Lambda 31}c_{\Lambda 33}s_{\Lambda 32}}&
{c_{\Lambda 33}c_{\Lambda 32}}\end {array}\right ]
\left [\begin {array}{c} {\Lambda_{13}}\\\noalign{\medskip}{\Lambda_{33}}
\\\noalign{\medskip}{{\rm Singlet}_{13}}
\end {array}\right ] \]
Here $\Lambda(??)$ is the unidentified spin-$3/2$ \rep{70}-plet $\Lambda$
state, orthogonal to $\Lambda(1690)$ and $\Lambda(1520)$. Although the
physical state has not been identified, we can make predictions for its decay
widths into the allowed \rep{56}-pion channels, provided that we make a
reasonable guess at its mass.  Finally, only one of the spin-$3/2$
$\Sigma$ states has been identified, and we parametrize it as follows:
\beq \Sigma(1775)= \left [\begin {array}{ccc}
c_{\Sigma 31} c_{\Sigma 32} &
s_{\Sigma 31} c_{\Sigma 32} &
s_{\Sigma 32} \end {array} \right ]
\left [\begin {array}{c} {\Sigma_{13}}\\\noalign{\medskip}{\Sigma_{33}}
\\\noalign{\medskip}{\Sigma^{\ast}_{13}}
\end {array}\right ] \eeq
For the spin-$1/2$ nucleon pure d-wave decays (for which only upper bounds are
known), we use the mixing angle obtained by fitting the s-wave decays, namely
$\theta_{N1} = 0.61$.

The coefficients of the operators $D$, $E$, $F$, and $G$ together with the six
new mixing angles, combine to give us ten parameters. With 25 decay fractions
to be fitted, there are 15 degrees of freedom. The best fit has
$\chi^2=36.0$. As with the s-wave decays, although the coefficients $d$, $e$,
$f$, and $g$ are reasonably constrained by the fitting procedure, there are
several solutions for the mixing angles, all of which have a value of $\chi^2$
close to the minimum value.  We obtain two solutions for the spin-$3/2$
nucleon mixing matrix, four solutions for spin-$3/2$ $\Lambda$ mixing,
and two solutions for the $\Sigma(1775)$ state.  All the solutions for the
three coefficients and the six mixing angles are tabulated in
Table~\ref{dparams}.  The corresponding spin-$3/2$ mixing matrices are found
in Tables~\ref{dnucmix}-\ref{dsigmix}.  The calculated decay fractions for
each set of parameters are presented in Tables~\ref{dnomix}-\ref{dsigma},
together with the corresponding experimental data.


\begin{table}[htbp]
\begin{center}
\begin{tabular}{lccc} \hline \hline
& $\Gamma_{\tot}^{(\xp)}(\MeV)$ &
\multicolumn{1}{c}{$\Gamma_{\tot}^{(\pred)}(\MeV)$} \\
$\Delta(1620)$ & $150 \pm 30$ & 138.8 \\ \hline
& $f_i^{(\xp)}(\%)$ & \multicolumn{1}{c}{$f_i^{(\pred)}(\%)$} \\
$\rightarrow \Delta \pi$ & $50.0 \pm 14.1$ & 41.8 \\ \hline \hline
& $\Gamma_{\tot}^{(\xp)}(\MeV)$ &
\multicolumn{1}{c}{$\Gamma_{\tot}^{(\pred)}(\MeV)$} \\
$\Delta(1700)$ & $300 \pm 100$ & 259.7 \\ \hline
& $f_i^{(\xp)}(\%)$ & \multicolumn{1}{c}{$f_i^{(\pred)}(\%)$} \\
$\rightarrow N \pi$ & $15.0 \pm 5.8$ & 12.0 \\ \hline \hline
& $\Gamma_{\tot}^{(\xp)}(\MeV)$ &
\multicolumn{1}{c}{$\Gamma_{\tot}^{(\pred)}(\MeV)$} \\
$N(1535)$ & $175 \pm 75$ & 186.5 \\ \hline
& $f_i^{(\xp)}(\%)$ & \multicolumn{1}{c}{$f_i^{(\pred)}(\%)$} \\
$\rightarrow \Delta \pi$ & $5.0 \pm 5.1$ & 0.4 \\ \hline \hline
& $\Gamma_{\tot}^{(\xp)}(\MeV)$ &
\multicolumn{1}{c}{$\Gamma_{\tot}^{(\pred)}(\MeV)$} \\
$N(1650)$ & $167.5 \pm 22.5$ & 173.1 \\ \hline
& $f_i^{(\xp)}(\%)$ & \multicolumn{1}{c}{$f_i^{(\pred)}(\%)$} \\
$\rightarrow \Delta \pi$ & $5.0 \pm 5.1$ & 9.0 \\ \hline \hline
\end{tabular}
\end{center}
\centerline{Table \ref{dnomix}: D-wave decays with no spin-$3/2$
mixing}
\end{table}

\begin{table}[htbp]
\begin{center}
\begin{tabular}{lccc}\hline\hline
& $\Gamma_{\tot}^{(\xp)}(\MeV)$ &
\multicolumn{1}{c}{$\Gamma_{\tot}^{(\pred)}(\MeV)$} \\
$N(1675)$ & $160 \pm 20$ & 158.0 \\ \hline
& $f_i^{(\xp)}(\%)$ & \multicolumn{1}{c}{$f_i^{(\pred)}(\%)$} \\
$\rightarrow N \pi$ & $45.0 \pm 10.3$ & 38.3 \\
$\rightarrow N \eta$ & $1.0 \pm 1.0$ & 2.1 \\
$\rightarrow \Lambda K$ & $0.1 \pm 0.1$ & 0.005 \\
$\rightarrow \Delta \pi$ & $55.0 \pm 12.1$ & 53.7 \\ \hline \hline
& $\Gamma_{\tot}^{(\xp)}(\MeV)$ &
\multicolumn{1}{c}{$\Gamma_{\tot}^{(\pred)}(\MeV)$} \\
$\Lambda(1830)$ & $85 \pm 25$ & 108.2 \\ \hline
& $f_i^{(\xp)}(\%)$ & \multicolumn{1}{c}{$f_i^{(\pred)}(\%)$} \\
$\rightarrow N \overline{K}$ & $6.5 \pm 3.7$ & 1.3 \\
$\rightarrow \Sigma \pi$ & $55.0 \pm 22.8$ & 83.2 \\ \hline \hline
& $\Gamma_{\tot}^{(\xp)}(\MeV)$ &
\multicolumn{1}{c}{$\Gamma_{\tot}^{(\pred)}(\MeV)$} \\
$\Sigma(1775)$ & $120 \pm 15$ & 124.5 \\ \hline
& $f_i^{(\xp)}(\%)$ & \multicolumn{1}{c}{$f_i^{(\pred)}(\%)$} \\
$\rightarrow N \overline{K}$ & $40.0 \pm 8.5$ & 17.3 \\
$\rightarrow \Lambda \pi$ & $17.0 \pm 4.5$ & 25.6 \\
$\rightarrow \Sigma \pi$ & $3.5 \pm 1.7$ & 3.4 \\
$\rightarrow \Sigma^{\ast} \pi$ & $10.0 \pm 2.8$ & 6.7 \\ \hline \hline
\end{tabular}
\caption{D-wave decays with no spin-$3/2$ mixing (continued)}
\label{dnomix}
\end{center}
\end{table}

\begin{table}[htbp]
\begin{center}
\begin{tabular}{lccc} \hline \hline
& & fit \#1 & fit \#2 \\ \hline \hline
& $\Gamma_{\tot}^{(\xp)}(\MeV)$ &
\multicolumn{2}{c}{$\Gamma_{\tot}^{(\pred)}(\MeV)$} \\
$N(1520)$ & $122.5 \pm 12.5$ & $128.0$ & $128.0$ \\ \hline
& $f_i^{(\xp)}(\%)$ & \multicolumn{2}{c}{$f_i^{(\pred)}(\%)$} \\
$\rightarrow N \pi$ & $55.0 \pm 12.1$ & 65.5 & 65.4 \\
$\rightarrow N \eta$ & $0.1 \pm 0.1$ & 0.07 & 0.08 \\ \hline \hline
& $\Gamma_{\tot}^{(\xp)}(\MeV)$ &
\multicolumn{2}{c}{$\Gamma_{\tot}^{(\pred)}(\MeV)$} \\
$N(1700)$ & $100 \pm 50$ & $101.3$ & $107.3$ \\ \hline
& $f_i^{(\xp)}(\%)$ & \multicolumn{2}{c}{$f_i^{(\pred)}(\%)$} \\
$\rightarrow N \pi$ & $10.0 \pm 5.4$ & 13.2 & 11.9 \\
$\rightarrow \Lambda K$ & $0.2 \pm 0.1$ & 0.09 & 0.03 \\ \hline \hline
\end{tabular}
\caption{D-wave decays for spin-$3/2$ nucleon initial states}
\label{dnucleon}
\end{center}
\end{table}

\begin{table}[htbp]
\begin{center}
\begin{tabular}{lccc} \hline \hline
& & fits \#1,3 & fits \#2,4 \\ \hline \hline
& $\Gamma_{\tot}^{(\xp)}(\MeV)$ &
\multicolumn{2}{c}{$\Gamma_{\tot}^{(\pred)}(\MeV)$} \\
$\Lambda(1690)$ & $60 \pm 10$ & $57.6$ & $55.2$ \\ \hline
& $f_i^{(\xp)}(\%)$ & \multicolumn{2}{c}{$f_i^{(\pred)}(\%)$} \\
$\rightarrow N \overline{K}$ & $25.0 \pm 7.1$ & 21.7 & 20.9 \\
$\rightarrow \Sigma \pi$ & $30.0 \pm 11.7$ & 30.3 & 24.9 \\ \hline \hline
& $\Gamma_{\tot}^{(\xp)}(\MeV)$ &
\multicolumn{2}{c}{$\Gamma_{\tot}^{(\pred)}(\MeV)$} \\
$\Lambda(1520)$ & $15.6 \pm 1.0$ & $15.2$ & $15.1$ \\ \hline
& $f_i^{(\xp)}(\%)$ & \multicolumn{2}{c}{$f_i^{(\pred)}(\%)$} \\
$\rightarrow N \overline{K}$ & $45.0 \pm 9.1$ & 17.9 & 14.4 \\
$\rightarrow \Sigma \pi$ & $42.0 \pm 8.5$ & 41.5 & 36.3 \\ \hline \hline
\end{tabular}
\caption{D-wave decays for spin-$3/2$ $\Lambda$ initial states}
\label{dlambda}
\end{center}
\end{table}

\begin{table}[htbp]
\begin{center}
\begin{tabular}{lccc} \hline \hline
& & fit \#1 & fit \#2 \\ \hline \hline
& $\Gamma_{\tot}^{(\xp)}(\MeV)$ &
\multicolumn{2}{c}{$\Gamma_{\tot}^{(\pred)}(\MeV)$} \\
$\Sigma(1670)$ & $60 \pm 20$ & $49.5$ & $7.3$ \\ \hline
& $f_i^{(\xp)}(\%)$ & \multicolumn{2}{c}{$f_i^{(\pred)}(\%)$} \\
$\rightarrow N \overline{K}$ & $10.0 \pm 3.6$ & 4.0 & 8.4 \\
$\rightarrow \Lambda \pi$ & $10.0 \pm 6.4$ & 11.6 & 10.8 \\
$\rightarrow \Sigma \pi$ & $45.0 \pm 17.5$ & 44.4 & 43.5 \\ \hline \hline
\end{tabular}
\caption{D-wave decays for spin-$3/2$ $\Sigma$ initial states}
\label{dsigma}
\end{center}
\end{table}

\begin{table}[htbp]
\begin{center}
\begin{tabular}{lcccc} \hline \hline
Parameter & value \\ \hline \hline
d & $0.203 \pm 0.011$ \\
e & $-0.015 \pm 0.004$ \\
f & $-0.029 \pm 0.008$ \\
g & $-0.002 \pm 0.005$ \\ \hline \hline
& fit \#1 & fit \#2 \\
$\theta_{N3}$ & $3.04 \pm 0.15$ & $2.60 \pm 0.16$ \\
$\Delta \chi^2$ & 0.00 & 1.27 \\ \hline \hline
& fit \#1 & fit \#2 & fit \#3 & fit \#4 \\
$\theta_{\Lambda 31}$ & $0.32 \pm 0.25$ & $1.04 \pm 0.18$ & $2.20 \pm 0.25$ &
$1.45 \pm 0.18$ \\
$\theta_{\Lambda 32}$ & $0.14 \pm 0.08$ & $2.61 \pm 0.10$ & $2.93 \pm 0.07$ &
$0.42 \pm 0.10$ \\
$\theta_{\Lambda 33}$ & $2.63 \pm 0.17$ & $0.45 \pm 0.15$ & $0.42 \pm 0.17$ &
$2.72 \pm 0.15$ \\
$\Delta \chi^2$ & 0.00 & 3.55 & 0.00 & 3.55 \\ \hline \hline
& fit \#1 & fit \#2 \\
$\theta_{\Sigma 31}$ & $2.14 \pm 0.37$ & $1.00 \pm 0.11$ \\
$\theta_{\Sigma 32}$ & $0.48 \pm 0.22$ & $0.76 \pm 0.34$ \\
$\Delta \chi^2$ & 0.00 & 4.03 \\ \hline \hline
\end{tabular}
\caption{Parameters from d-wave fit}
\label{dparams}
\end{center}
\end{table}

\begin{table}[htbp]
\[ M_{N3}^{(\fit \#1)} = \left [\begin {array}{cc}
-0.99 \pm 0.02 & 0.10 \pm 0.15
\\\noalign{\medskip}
-0.10 \pm 0.15 & -0.99 \pm 0.02
\end {array}
\right ] \]

\[ M_{N3}^{(\fit \#2)} = \left [\begin {array}{cc}
-0.85 \pm 0.08 & 0.52 \pm 0.14
\\\noalign{\medskip}
-0.52 \pm 0.14 & -0.85 \pm 0.08
\end {array}
\right ] \]
\caption{Spin-$3/2$ nucleon mixing matrices}
\label{dnucmix}
\end{table}

\begin{table}[htbp]
\[ M_{\Lambda 3}^{(\fit \#1)} = \left [\begin {array}{ccc}
0.94 \pm 0.07 & 0.31 \pm 0.24 & 0.14 \pm 0.08
\\\noalign{\medskip}
0.20 \pm 0.24 & -0.85 \pm 0.10 & 0.48 \pm 0.15
\\\noalign{\medskip}
0.27 \pm 0.09 & -0.43 \pm 0.15 & -0.86 \pm 0.08
\end {array}\right ] \]

\[ M_{\Lambda 3}^{(\fit \#2)} = \left [\begin {array}{ccc}
-0.44 \pm 0.15 & -0.74 \pm 0.08 & 0.51 \pm 0.09
\\\noalign{\medskip}
-0.89 \pm 0.07 & 0.27 \pm 0.18 & -0.37 \pm 0.13
\\\noalign{\medskip}
0.14 \pm 0.16 & -0.61 \pm 0.06 & -0.78 \pm 0.06
\end {array}\right ] \]

\[ M_{\Lambda 3}^{(\fit \#3)} = \left [\begin {array}{ccc}
0.58 \pm 0.19 & -0.79 \pm 0.15 & 0.21 \pm 0.07
\\\noalign{\medskip}
-0.69 \pm 0.15 & -0.61 \pm 0.19 & -0.40 \pm 0.16
\\\noalign{\medskip}
0.44 \pm 0.12 & 0.09 \pm 0.14 & -0.89 \pm 0.06
\end {array}\right ] \]

\[ M_{\Lambda 3}^{(\fit \#4)} = \left [\begin {array}{ccc}
0.11 \pm 0.16 & 0.91 \pm 0.05 & 0.41 \pm 0.09
\\\noalign{\medskip}
0.89 \pm 0.08 & -0.28 \pm 0.18 & 0.37 \pm 0.13
\\\noalign{\medskip}
0.45 \pm 0.14 & 0.32 \pm 0.11 & -0.83 \pm 0.05
\end {array}\right ] \]
\caption{Spin-$3/2$ $\Lambda$ mixing matrices}
\label{dlammix}
\end{table}

\begin{table}[htbp]
\[ M_{\Sigma 3}^{(\fit \#1)} = \left [\begin {array}{ccc}
-0.48 \pm 0.27 & 0.75 \pm 0.22 &  0.46 \pm 0.19
\end {array}\right ] \]

\[ M_{\Sigma 3}^{(\fit \#2)} = \left [\begin {array}{ccc}
0.39 \pm 0.10 & 0.61 \pm 0.23 & 0.69 \pm 0.24
\end {array}\right ] \]
\caption{Spin-$3/2$ $\Sigma$ mixing matrices}
\label{dsigmix}
\end{table}

We see from Table~\ref{dparams} that the coefficients $e$, $f$, and $g$ are
strongly suppressed relative to $d$, which is consistent with what we
found for the s-wave decays.  Comparing predicted and experimental branching
fractions, we see that a large part of the total $\chi^2$ comes from the
$\Lambda(1520) \rightarrow N \overline{K}$ decay. $\Sigma(1775) \rightarrow
N\overline{K}$ also seems strongly enhanced relative to our predictions.

\section{Decay predictions} \label{app:pred}

In this appendix we predict the partial widths for all of the
remaining kinematically allowed one-pion decays.  We display a different
set of predictions corresponding to each of the fits presented in
Appendix~\ref{app:fits}.

While it was more convenient for us to work with branching fractions in the
previous section, here we present our results directly in terms of partial
widths.  The errors we present for these predictions are a combination of the
uncertainties in the parameters given in Tables~\ref{spars} and \ref{dparams},
and the uncertainties in the masses of the initial states.  The latter have a
large effect on our predictions for the decays that are near threshold, due to
the momentum dependence of the squared amplitudes.  For decays very near
threshold, we are able to obtain only an upper bound for the partial width.
In Tables~\ref{prednomix}-\ref{predsig3}, we list the decay predictions, the
total decay widths $\Gamma_{\rm tot}^{\rm (\pred)}$ given in Appendix
{}~\ref{app:fits}, and the experimentally measured total widths
$\Gamma_{\rm tot}^{\rm (\xp)}$.


\begin{table}[htbp]
\begin{center}
\begin{tabular}{lcc} \hline \hline
& $\Gamma_{\tot}^{(\xp)}(\MeV)$ & $\Gamma_{\tot}^{(\pred)}(\MeV)$ \\
$\Delta(1700)$ & $300 \pm 100 $ & $260$ \\ \hline
& & $\Gamma_{\i}^{(\pred)}{(\MeV)}$ \\
$\rightarrow \Delta \pi$ & & $271 \pm 126$ \\
\quad s-wave & & $241 \pm 117$ \\
\quad d-wave & & $30 \pm 18$ \\
$\rightarrow \Sigma K$ & & $<0.25$ \\  \hline \hline
& $\Gamma_{\tot}^{(\xp)}(\MeV)$ & $\Gamma_{\tot}^{(\pred)}(\MeV)$ \\
$\Lambda(1830)$ & $85 \pm 25$ & $108$ \\ \hline
& & $\Gamma_{\i}^{(\pred)}(\MeV)$ \\
$\rightarrow \Lambda \eta$ & & $4.9 \pm 1.0$ \\
$\rightarrow \Xi K$ & & $<0.01$ \\  \hline \hline
& $\Gamma_{\tot}^{(\xp)}(\MeV)$ & $\Gamma_{\tot}^{(\pred)}(\MeV)$ \\
$\Sigma(1775)$ & $120 \pm 15$ & $125$ \\ \hline
& & $\Gamma_{\i}^{(\pred)}(\MeV)$ \\
$\rightarrow \Sigma \eta$ & & $0.12 \pm 0.04$ \\
$\rightarrow \Delta \overline{K}$ & & $0.85 \pm 0.23$ \\  \hline \hline
\end{tabular}
\caption{Predictions involving no mixing angles}
\label{prednomix}
\end{center}
\end{table}

\begin{table}[htbp]
\begin{center}
\begin{tabular}{lccc} \hline \hline
& & fit \#1 & fit \#2 \\ \hline \hline
& $\Gamma_{\tot}^{(\xp)}(\MeV)$ &
\multicolumn{2}{c}{$\Gamma_{\tot}^{(\pred)}(\MeV)$} \\
$N(1520)$ & $122.5 \pm 12.5$ & $128.0$ & $128.0$ \\ \hline
& &\multicolumn{2}{c}{$\Gamma_{\i}^{(\pred)}(\MeV)$} \\
$\rightarrow \Delta \pi$ & & $18.0 \pm 4.4$ & $9.8 \pm 1.9$ \\
\quad s-wave & & {$10.6 \pm 3.4$}
& {$0.03 \pm 0.01$} \\
\quad d-wave & & {$7.5 \pm 2.5$}
& {$9.8 \pm 1.9$} \\ \hline \hline
& $\Gamma_{\tot}^{(\xp)}(\MeV)$ &
\multicolumn{2}{c}{$\Gamma_{\tot}^{(\pred)}(\MeV)$} \\
$N(1700)$ & $100 \pm 50 $ & $101.3$ & $107.3$ \\ \hline
& &\multicolumn{2}{c}{$\Gamma_{\i}^{(\pred)}(\MeV)$} \\
$\rightarrow \Delta \pi$ & & $180 \pm 74$ & $189 \pm 76$ \\
\quad s-wave & & {$151 \pm 61$}
& {$188 \pm 76$} \\
\quad d-wave & & {$29 \pm 27$}
& {$<5$} \\
$\rightarrow N \eta$ & & $1.5 \pm 1.2$ & $<0.2$ \\
$\rightarrow \Sigma K$ & & $<0.03$ & $<0.004$ \\ \hline \hline
\end{tabular}
\caption{Predictions for decays of spin-3/2 N initial states}
\label{prednuc3}
\end{center}
\end{table}

\begin{table}[htbp]
\begin{center}
\begin{tabular}{lccccc} \hline \hline
& & fit \#1 & fit \#2 & fit \#3 & fit \#4 \\ \hline
& $\Gamma_{\tot}^{(\xp)}(\MeV)$
&\multicolumn{4}{c}{$\Gamma_{\tot}^{(\pred)}(\MeV)$} \\
$\Lambda(1670)$ & $37.5 \pm 12.5$ & $38.1$ & $38.0$ & $32.3$
& $38.0$ \\ \hline
& &\multicolumn{4}{c}{$\Gamma_{\i}^{(\pred)}(\MeV)$} \\
$\rightarrow \Sigma^* \pi$ & & $0.72 \pm 0.36$ & $0.034 \pm 0.017$
& $<0.03$ & $0.034 \pm 0.017$ \\  \hline \hline
& & fit \#5 & fit \#6 & fit \#7 & fit \#8 \\  \hline
& &\multicolumn{4}{c}{$\Gamma_{\tot}^{(\pred)}(\MeV)$} \\
$\Lambda(1670)$ & & $32.3$ & $38.1$ & $31.5$
& $31.5$ \\ \hline
& &\multicolumn{4}{c}{$\Gamma_{\i}^{(\pred)}(\MeV)$} \\
$\rightarrow \Sigma^* \pi$ & & $<0.03$
& $0.72 \pm 0.36$ & $0.23 \pm 0.11$ & $0.23 \pm 0.12$ \\
\hline \hline
& & fit \#1 & fit \#2 & fit \#3 & fit \#4 \\ \hline
& $\Gamma_{\tot}^{(\xp)}(\MeV)$ &
\multicolumn{4}{c}{$\Gamma_{\tot}^{(\pred)}(\MeV)$} \\
$\Lambda(1800)$ & $300 \pm 100$ & $300.7$ & $300.6$ & $299.0$
& $300.6$ \\ \hline
& &\multicolumn{4}{c}{$\Gamma_{\i}^{(\pred)}(\MeV)$} \\
$\rightarrow \Sigma \pi$ & & $187 \pm 116$ & $170 \pm 101$
& $191 \pm 109$ & $148 \pm 59$ \\
$\rightarrow \Sigma^* \pi$ & & $0.53 \pm 0.44$ & $1.5 \pm 1.2$
& $1.5 \pm 1.2$ & $15 \pm 12$ \\
$\rightarrow \Lambda \eta$ & & $18 \pm 14$ & $<0.25$
& $<4$ & $15 \pm 9$ \\ \hline \hline
& & fit \#5 & fit \#6 & fit \#7 & fit \#8 \\ \hline
& &\multicolumn{4}{c}{$\Gamma_{\tot}^{(\pred)}(\MeV)$} \\
$\Lambda(1800)$ & & $299.0$ & $300.7$ & $299.7$
& $299.7$ \\ \hline
& &\multicolumn{4}{c}{$\Gamma_{\i}^{(\pred)}(\MeV)$} \\
$\rightarrow \Sigma \pi$ & & $130 \pm 59$
& $125 \pm 57$ & $149 \pm 61$ & $175 \pm 122$ \\
$\rightarrow \Sigma^* \pi$ & & $15 \pm 12$
& $9.3 \pm 7.8$ & $12 \pm 10$ & $0.33 \pm 0.28$ \\
$\rightarrow \Lambda \eta$ & & $24 \pm 13$
& $<2$ & $<4$ & $23 \pm 15$ \\
\hline \hline
\end{tabular}
\caption{Predictions for decays of spin-1/2 $\Lambda$ initial states}
\label{predlam1}
\end{center}
\end{table}

\begin{table}[htbp]
\begin{center}
\begin{tabular}{lccccc} \hline \hline
& & fit \#1 & fit \#2 & fit \#3 & fit \#4 \\ \hline \hline
& $\Gamma_{\tot}^{(\xp)}(\MeV)$
&\multicolumn{4}{c}{$\Gamma_{\tot}^{(\pred)}(\MeV)$} \\
$\Lambda(1690)$ & $60 \pm 10$ & $57.6$ & $55.2$ & $57.6$
& $55.2$ \\ \hline
& &\multicolumn{4}{c}{$\Gamma_{\i}^{(\pred)}(\MeV)$} \\
$\rightarrow \Lambda \eta$ & & $<0.01$ & $<0.1$
& $<0.001$ & $<0.04$ \\
$\rightarrow \Sigma^* \pi$ & & $32 \pm 10$ & $36 \pm 12$
& $15.8 \pm 2.9$ & $36 \pm 11$ \\
\quad s-wave & & {$28.5 \pm 9.1$} & {$36 \pm 12$}
& {$6.3 \pm 2.0$} & {$33 \pm 11$} \\
\quad d-wave & & {$3.9 \pm 3.1$} & {$<0.4$}
& {$9.5 \pm 2.1$} & {$2.1 \pm 1.8$} \\
\hline \hline
& $\Gamma_{\tot}^{(\xp)}(\MeV)$
&\multicolumn{4}{c}{$\Gamma_{\tot}^{(\pred)}(\MeV)$} \\
$\Lambda(??)$ & -- & -- & -- &  -- &
-- \\ \hline
& &\multicolumn{4}{c}{$\Gamma_{\i}^{(\pred)}(\MeV)$} \\
$\rightarrow N \overline{K}$ & & $<3$ & $38 \pm 29$
& $<3$ & $38 \pm 29$ \\
$\rightarrow \Sigma \pi$ & & $105 \pm 69$ & $140 \pm 65$
& $105 \pm 69$ & $139 \pm 65$ \\
$\rightarrow \Lambda \eta$ & & $2.6 \pm 3.3$ & $<0.4$
& $<2.5$ & $<0.4$ \\
$\rightarrow \Sigma^* \pi$ & & $97 \pm 46$ & $85 \pm 43$
& $120 \pm 49$ & $85 \pm 43$ \\
\quad s-wave & & $55 \pm 22$ & $9.3 \pm 3.7$
& $116 \pm 47$ & $8.6 \pm 3.5$ \\
\quad d-wave & & $42 \pm 32$ & $75 \pm 41$
& $<15$ & $76 \pm 41$ \\
\hline \hline
\end{tabular}
\caption{Predictions for decays of spin-3/2 $\Lambda$ initial states}
\label{predlam3}
\end{center}
\end{table}

\begin{table}[htbp]
\begin{center}
\begin{tabular}{lccccc} \hline \hline
& & fit \#1 & fit \#2 & fit \#3 & fit \#4 \\ \hline \hline
& $\Gamma_{\tot}^{(\xp)}(\MeV)$
&\multicolumn{4}{c}{$\Gamma_{\tot}^{(\pred)}(\MeV)$} \\
$\Sigma(1750)$ & $110 \pm 50$ & $109.7$ & $109.7$ & $110.1$
& $110.1$ \\ \hline
& &\multicolumn{4}{c}{$\Gamma_{\i}^{(\pred)}(\MeV)$} \\
$\rightarrow \Lambda \pi$ & & $<7$ & $43 \pm 22$
& $<20$ & $49 \pm 12$ \\
$\rightarrow \Sigma^* \pi$ & & $25 \pm 17$ & $<0.7$
& $22 \pm 15$ & $2.1 \pm 1.6$ \\
$\rightarrow \Delta \overline{K}$ & & $<1.4$ & $<0.9$
& $<2.5$ & $<1.7$ \\
\hline \hline
\end{tabular}
\caption{Predictions for decays of spin-1/2 $\Sigma$ initial states}
\label{predsig1}
\end{center}
\end{table}

\begin{table}[htbp]
\begin{center}
\begin{tabular}{lccc} \hline \hline
& & fit \#1 & fit \#2 \\ \hline \hline
& $\Gamma_{\tot}^{(\xp)}(\MeV)$
&\multicolumn{2}{c}{$\Gamma_{\tot}^{(\pred)}(\MeV)$} \\
$\Sigma(1670)$ & $60 \pm 20$ & $49.5$ & $7.3$ \\ \hline
& &\multicolumn{2}{c}{$\Gamma_{\i}^{(\pred)}(\MeV)$} \\
$\rightarrow \Sigma^* \pi$ & & $15.8 \pm 4.8$ & $41 \pm 12$ \\
\quad s-wave & & $15.5 \pm 4.6$ & $40 \pm 12$ \\
\quad d-wave & & $0.27 \pm 0.35$ & $0.57 \pm 0.61$ \\
\hline \hline
\end{tabular}
\caption{Predictions for decays of spin-3/2 $\Sigma$ initial states}
\label{predsig3}
\end{center}
\end{table}

Among our predictions are six decays that can proceed through both the
s- and d-wave channels.  (We will refer to these as s+d-wave decays.)
Three of these have been measured reasonably well, while the others are
either poorly known or unobserved.  We have chosen not to include the former
three in our fits in Appendix~\ref{app:fits}, to simplify our analysis.  In
principle, a proper treatment would require fitting the pure s-wave, the pure
d-wave, and the s+d-wave decays simultaneously.  Instead we simply check in
this section that the predictions for the three measured s+d-wave decays are
in reasonable agreement with the experimental results.

\centerline{{\em Decays involving no mixing angles}}

We first consider predictions of the partial decay widths that do not
involve mixing angles.   The unmixed initial states are the spin-1/2
$\Delta(1620)$, the spin-3/2 $\Delta(1700)$, and the spin-5/2
$N(1675)$, $\Lambda(1830)$, and $\Sigma(1775)$. The kinematically
allowed decays that we have not already considered in Appendix~\ref{app:fits}
are listed in Table~\ref{prednomix}.  The
$\Delta(1700) \rightarrow \Delta \pi$ is one of the s+d-wave decays that
have been adequately measured.  Our prediction of $271 \pm 126$ MeV is
compatible with the RPP's value of $(.45\pm 0.1)\times(300 \pm 100)$
MeV if we take the partial width to lie at the lower end of the predicted
range. Analysis of the dependence of our prediction on the various parameters
suggests that the $c$ parameter should take its value at the bottom of
the range given in Table~\ref{spars}.

\centerline{{\em Nucleon decays}}

All the kinematically allowed spin-1/2 $N$ decays have been included in
Appendix~\ref{app:fits}. Predictions for the spin-3/2 nucleons are shown in
Table~\ref{prednuc3}. The $N(1520) \rightarrow \Delta
\pi$ is another of the three known s+d-wave decays. The RPP's
value $(.22 \pm .08)\times(122 \pm 13)$ MeV for its partial width
strongly favors fit \#1. Furthermore, the RPP's partial wave analysis of this
decay is consistent with fit \#1, but incompatible with fit \#2.
Therefore we conclude that fit \#1 has the correct mixing angle.

The $N(1700) \rightarrow \Delta \pi$ is the third of the known
s+d-wave decays. Our fit~\#1 prediction of $180 \pm 74$ MeV is
large compared to the RPP's value of $(.38 \pm .32)\times(100 \pm 50)$ MeV.
However, if we adopt a value of the $N(1700)$ full width that is just
within the RPP's upper limit, while also using the smallest
prediction for the $N(1700) \rightarrow \Delta \pi$ width
consistent with our range of error, we obtain a branching
fraction that is in reasonable agreement with experiment.
This increase in the total decay width still allows good fits for
the $N(1700) \rightarrow N \pi$ and $N(1700) \rightarrow \Lambda K$ decay
fractions (see Table~\ref{dnucleon}).  If we had included
the $N(1700) \rightarrow \Delta \pi$ decay in the fits in
Appendix~\ref{app:fits}, the only substantial change would have
been an increase in the $N(1700)$ predicted full width.

\centerline{{\em Lambda decays}}

There are four allowed spin-1/2 $\Lambda$ decays, listed in
Table~\ref{predlam1}.  The one measurement in the RPP for the
$\Lambda (1670) \rightarrow \Sigma^* \pi$ width, 6 $\pm$ 3 MeV, is
somewhat larger than our predictions, and favors fits \#1 and \#6.
Of the six spin-3/2 $\Lambda$ decays shown in Table~\ref{predlam3}, four
involve the unobserved $\Lambda$ state which is orthogonal to the
$\Lambda(1520)$ and $\Lambda(1690)$.  To compute the decay widths of
the unobserved state, we made a reasonable guess at its mass
based on the nucleon-lambda splitting found in other multiplets.
The mass we adopted was $1850 \pm 50$ MeV.  The four widths involving
the unobserved state  have large errors in part because there are no known
decays to fit the mixing angles more accurately, and in part because there
is a large uncertainty in the mass.

\centerline{{\em Sigma decays}}

Three spin-1/2 $\Sigma$ decays are listed in Table~\ref{predsig1}, and one
spin-3/2 $\Sigma$ decay in Table~\ref{predsig3}.  Since we know only two out
of the three $\theta_{\Sigma}$ mixing angles for both the spin-1/2 and
spin-3/2 $\Sigma$'s (see Tables~\ref{ssigmix} and
\ref{dsigmix}), we know the orientation in the 3-dimensional mixing space for
only one spin-1/2 $\Sigma$ (the $\Sigma(1750)$) and only one spin-3/2
$\Sigma$ (the $\Sigma(1670)$). We therefore can not make any
predictions concerning the spin-1/2 $\Sigma(1620)$, or the
three unobserved $\Sigma$ states.

In Table~\ref{predsig3}, fit \#2 for the spin-3/2 $\Sigma$'s does not
appear to be acceptable; the $\Sigma(1670) \rightarrow \Sigma^* \pi$
branching fraction, combined with the branching fractions in
Table~\ref{dsigma}, yields a sum  greater than unity. Fit \#1, however, is
consistent with the data available in the RPP.


\input prepictex
\input pictex
\input postpictex

\newsavebox{\sru}
\savebox{\sru}{\beginpicture
\setcoordinatesystem units <\unitlength,\unitlength>
\circulararc 90 degrees from 0 0 center at 0 7.5
\circulararc -90 degrees from 10 0 center at 10 7.5
\circulararc 180 degrees from 7.5 7.5 center at 5 7.5
\endpicture}
\newcommand{\springru}[1]{\multiput(0,0)(10,0){#1}{\usebox{\sru}}}
\newsavebox{\srd}
\savebox{\srd}{\beginpicture
\setcoordinatesystem units <\unitlength,\unitlength>
\circulararc -90 degrees from 0 0 center at 0 -7.5
\circulararc 90 degrees from 10 0 center at 10 -7.5
\circulararc -180 degrees from 7.5 -7.5 center at 5 -7.5
\endpicture}
\newcommand{\springrd}[1]{\multiput(0,0)(10,0){#1}{\usebox{\srd}}}
\newsavebox{\sdr}
\savebox{\sdr}{\beginpicture
\setcoordinatesystem units <\unitlength,\unitlength>
\circulararc 90 degrees from 0 0 center at 7.5 0
\circulararc -90 degrees from 0 -10 center at 7.5 -10
\circulararc 180 degrees from 7.5 -7.5 center at 7.5 -5
\endpicture}
\newcommand{\springdr}[1]{\multiput(0,0)(0,-10){#1}{\usebox{\sdr}}}
\newsavebox{\sdl}
\savebox{\sdl}{\beginpicture
\setcoordinatesystem units <\unitlength,\unitlength>
\circulararc -90 degrees from 0 0 center at -7.5 0
\circulararc 90 degrees from 0 -10 center at -7.5 -10
\circulararc -180 degrees from -7.5 -7.5 center at -7.5 -5
\endpicture}
\newcommand{\springdl}[1]{\multiput(0,0)(0,-10){#1}{\usebox{\sdl}}}

\def\tarrow{\arrow <5pt> [.3,.6]}
\def\square{\beginpicture\setcoordinatesystem units <\tdim,\tdim>
\putrule from -5 -5 to -5 5
\putrule from 5 -5 to 5 5
\putrule from -5 -5 to 5 -5
\putrule from -5 5 to 5 5
\endpicture}

\newpage

\section*{Figures}

\begin{figure}[htbp]
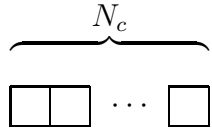

$$\beginpicture\setcoordinatesystem units <\tdim,\tdim>
\put {\square} at 0 0
\put {\square} at 10 0
\put {\square} at 40 0
\put {$\cdots$} at 25 0
\put {$\displaystyle\overbrace{\beginpicture\setcoordinatesystem units
<\tdim,\tdim>
\linethickness=0pt\putrule from 0 0 to 50 0
\endpicture}^{\displaystyle N_c}$} at 20 20
\endpicture$$\caption{\label{youngsym} Young tableaux for the spin-flavor
representation of the ground-state baryons for large $N_c$.}\end{figure}

\begin{figure}[htbp]
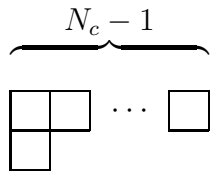

$$\beginpicture\setcoordinatesystem units <\tdim,\tdim>
\put {\square} at 0 0
\put {\square} at 0 -10
\put {\square} at 10 0
\put {\square} at 40 0
\put {$\cdots$} at 25 0
\put {$\displaystyle\overbrace{\beginpicture\setcoordinatesystem units
<\tdim,\tdim>
\linethickness=0pt\putrule from 0 0 to 50 0
\endpicture}^{\displaystyle N_c-1}$} at 20 20
\endpicture$$\caption{\label{youngmix1} Young tableaux for the spin-flavor
representation of the first excited $\ell=1$ baryons for large
$N_c$.}\end{figure}

\begin{figure}[htbp]
$$\begin{picture}(200,120)(-100,-40)
\thicklines
\multiput(0,0)(0,-5){8}{\line(0,-1){3}}
\put(-100,0){\line(1,0){200}}
\put(-100,40){\line(1,0){200}}
\put(-100,80){\line(1,0){200}}
\put(-30,40){\springdl{4}}
\put(30,80){\springdr{8}}
\put(-30,20){\springru{6}}
\end{picture}$$
\caption{\label{diagram} Feynman graph for a multiquark operator contributing
to the pion-baryon coupling in large $N_c$.}
\end{figure}
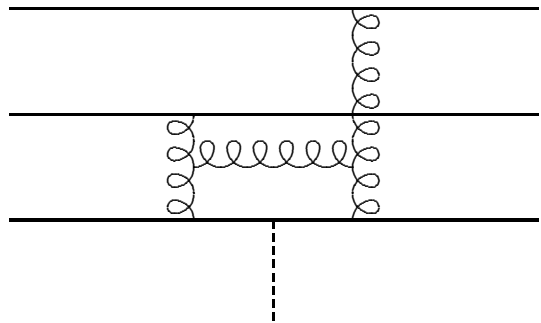

\end{document}